\documentclass[noshowpacs,noshowkeys,floatfix,amsmath,amssymb, aps,pra,reprint,superscriptaddress]{revtex4-1}
\usepackage{times}
\usepackage{graphicx}
\usepackage{dcolumn}
\usepackage{bm}
\usepackage{mathtools}
\usepackage[UKenglish]{babel}
\usepackage{physics}
\usepackage[shortlabels]{enumitem}
\usepackage{mathrsfs}
\usepackage{xfrac}
\usepackage{hyperref}
\usepackage{makecell}
\usepackage[all]{hypcap}
\usepackage{bbold}
\usepackage{color}
\usepackage{soul} 
\usepackage{epstopdf}

\DeclareMathOperator{\sgn}{sgn}

\begin{document}
\preprint{APS/123-QED}
\title{Analytical description of quantum emission in optical analogues to gravity}

\author{Maxime J Jacquet}
 \email{maxime.jacquet@lkb.upmc.fr}
 \affiliation{Laboratoire Kastler Brossel, Sorbonne Universit\'{e}, CNRS, ENS-Universit\'{e} PSL, Coll\`{e}ge de France, Paris 75005, France}
 \affiliation{Faculty of Physics, University of Vienna, Boltzmanngasse 5, Vienna A-1090, Austria.}
\author{Friedrich K\"{o}nig}%
 \email{fewk@st-andrews.ac.uk}
\affiliation{School of Physics and Astronomy, SUPA, University of St. Andrews, North Haugh, St. Andrews, KY16 9SS, United Kingdom}%
\date{\today}

\begin{abstract}
We consider a moving refractive index perturbation in an optical medium as an optical analogue to waves under the influence of gravity.
We describe the dielectric medium by the Lagrangian of the Hopfield model.
We supplement the field theory in curved spacetime for this model to solve the scattering problem for all modes and frequencies analytically.
Because of dispersion, the kinematic scenario of the field modes may contain optical event horizons for some frequencies.
We calculate the spectra of spontaneous emission in the frame co-moving with the perturbation and in the laboratory frame.
We also calculate the spectrally-resolved photon number correlations in either frame.
The emitted multimode field comes in different types depending on the presence of horizons.
We show that these types are robust against changes in the system parameters and thus are genuine features of optical and non-optical analogues.
These methods and findings pave the way to new observations of analogue gravity in dispersive systems.
\end{abstract}
\maketitle

\section{Introduction}

A number of classical and semi-classical features of gravity can be reproduced in the laboratory: it is possible to create effectively curved spacetimes for waves in media \cite{unruh_experimental_1981} and, in particular, event horizons.
These horizons scatter waves and are predicted to spontaneously emit quanta by the Hawking effect \cite{hawking_black_1974,hawking_particle_1975}.
Recently, experiments in many different `fluid' systems, such as Bose-Einstein condensates \cite{lahav_realization_2010,munoz_de_nova_observation_2019}, water \cite{rousseaux_observation_2008,weinfurtner_measurement_2011,Rousseaux_PRL_2016,torres_rotational_2017,euve_wormholes_2017,euve_scattering_2020} and  polariton microcavities \cite{nguyen_acoustic_2015} have demonstrated horizons and studied the behaviour of waves in their vicinity. 
New developments in quantum fluids of light \cite{michel_superfluid_2018,Fontaine_disprel_2018,jacquet_fluids_2020} may soon enable novel analogue gravity experiments.

As in the astrophysical case \cite{hawking_particle_1975}, the paired spontaneous emission in analogues results from mixing of field modes of positive and negative Klein-Gordon norm at the horizon \cite{unruh_experimental_1981}.
Thus, pairs from horizons have been extensively studied for fluid systems, in which their nonseparability in various dispersive regimes has been investigated \cite{Busch_entanglementHR_2014,Busch_entanglementfluid_2014,finazzi_entangled_2014,boiron_quantum_2015,michel_phonon_2016,weinfurtner-BEC-HR-2016}, and are considered an unmistakable signature of the Hawking effect \cite{schutzhold-unruh-comment-2011,finke_observation_2016,fabbri_momentum_2018,coutant_low-frequency_2018-1,coutant_low-frequency_2018,isoard_quantum_2020}.

It is also possible to create effectively curved spacetimes for light in dispersive media. 
As an optical pulse moves through the dispersive medium, the pulse intensity locally raises the refractive index $n$ of the medium by the optical Kerr effect, creating a moving refractive index front (RIF).
Light under the pulse is slowed by the increased index, \textit{i.e.}, light of some frequencies will be slowed below the pulse speed and captured into the RIF.
This is in analogy with the kinematics of waves at a black-hole event horizon \cite{visser_acoustic_1998,leonhardt_laboratory_2002,schutzhold_EMW_2005,gorbach_light_2007,philbin_fiber-optical_2008, Hill_soliton-trapping_2009,faccio_analogue_2010,choudhary_efficient_2012,Jacquet_book_2018,drori_observation_2019}.
At other frequencies, however, the kinematics of light waves are different \cite{Jacquet_quantum_2015,jacquet_influence_2020}, recreating wave motion at a white-hole event horizon or two-way motion towards and away from the RIF.
Emission in all these kinematic scenarios occur simultaneously.

\begin{figure}[t]
\centering
\includegraphics[width=.6\columnwidth]{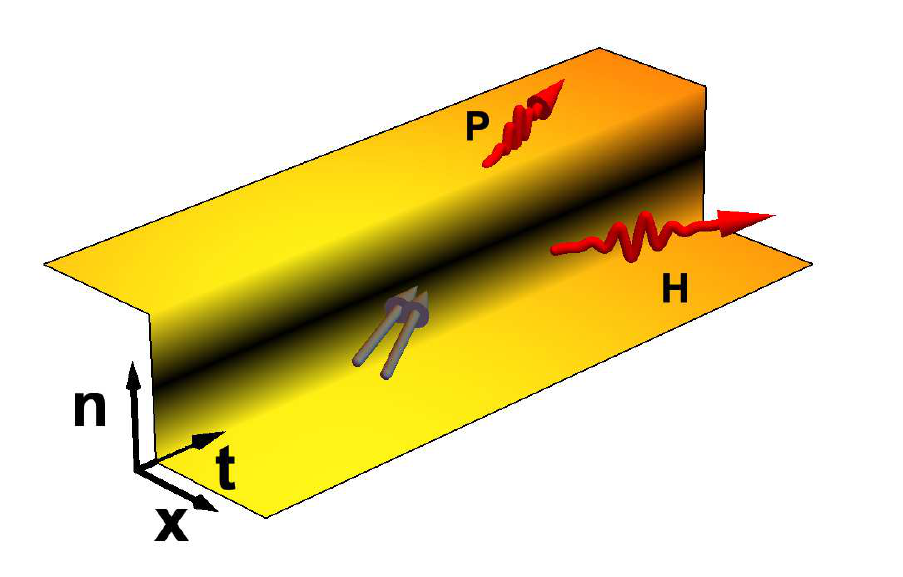}
\caption[RIF as analogue black hole horizon]{Schematic of the Hawking effect at the refractive index front (RIF). 
The RIF separates two homogeneous index regions on the left ($x<0$) and on the right ($x>0$) of a dielectric boundary of height $\delta n$.
The stationary process is shown in the frame co-moving with the RIF in the dispersive dielectric.
Incoming vacuum modes (grey) scatter on the boundary at $x=0$, which yields the spontaneous emission of entangled photons pairs.}\label{fig:HRstep}
\end{figure}

The RIF may be time independent in the moving frame: the system is stationary and can be described efficiently by the scattering matrix.
Possible analytical methods to arrive at the scattering matrix broadly fall under two categories, depending on the profile of the RIF considered: for smooth RIFs \cite{belgiorno_hawking_2015,linder_derivation_2016}, the JWKB approximation is used to calculate the asymptotic states (far from the RIF) that are matched with the near-RIF expansion (which is instead treated in Fourier space), whereas no approximations are needed in the case of a step-like RIF \cite{finazzi_quantum_2013}.
Although the step is physically hard to implement, because the particles are emitted from a single point, this simple optical system allows us to contrast the quantum emission in the different kinematic scenarios highlighted above.
The output fields will be further modified by gradual horizons, two-horizon interactions and other nonlinear effects such as optical \v{C}erenkov radiation. 
Studying the RIF allows us to identify which quantum effects are connected with a single horizon.
The approach of \cite{finazzi_quantum_2013} needs to be comprehensively generalised to account for the possible change of kinematic scenarios, and to include a treatment of evanescent waves in these scenarios as well as a derivation of the full scattering matrix.
From the scattering matrix follow the output quantum fields for all modes.
In particular, we are interested in key output observables --- such as photon-number spectra and correlations --- which are of principal importance in any theoretical or experimental study of analogue gravity.
The calculation of the photon-number correlations is challenging because it requires accounting for the dispersion and the bandwidth of the modes.

In this paper, we supplement and extend the theoretical description of the spontaneous emission at a step-like RIF of \cite{finazzi_quantum_2013}, as shown in Fig.\ref{fig:HRstep}.
We present an analytical method to calculate the scattering matrix in all kinematic scenarios in a  dispersive dielectric featuring intra- and inter-branch scattering.
Moreover, we demonstrate the calculation of key quantities measurable in the medium rest frame, \textit{viz}. the photon emission spectral density and the spectrally-resolved photon-number correlations, as they are recorded with a variable bandwidth detector.
We find pertinent features when modifying the parameters of the system.

In section \ref{sec:anagrav}, we review all possible kinematic scenarios for waves at the RIF and explain their relevance in the anlogy with gravity.
Section \ref{sec:analyticsmethods} derives the photon flux in different frames as well as the spectral correlations detected with variable bandwidth detectors in terms of the scattering matrix.
Section \ref{sec:analyticsSmethods} derives the scattering matrix for all modes and frequencies from the field matching conditions at the RIF.
In section \ref{sec:CS} we use our method to compute observables in bulk fused silica.
We calculate key observable quantities, namely the spectral density of spontaneous emission and the spectral correlations, in the rest frame of the medium and discuss their dependence on the RIF height and velocity.
We also briefly discuss desired medium properties.

\section{Kinematic scenarios}\label{sec:anagrav}
Our study is based on the consideration of the step-like geometry of a RIF that propagates at constant speed $u$ in the positive $X$-direction in the laboratory frame.
This RIF is illustrated in Figure \ref{fig:HRstep} in co-moving frame coordinates $x$ and $t$.
In Appendix \ref{app:FT} we review and supplement the field theory for light in an inhomogeneous dispersive dielectric.
There we find the modes of the homogeneous medium and construct the `global modes' (GMs) of the inhomogeneous system (including the step).
Finally we quantise the field which is represented by annihilation and creation operators of the GMs.
In this section, we review the kinematics of light waves at different frequencies \cite{Jacquet_quantum_2015} and explain the analogy with gravity.

\begin{figure}[t]
\centering
\includegraphics[width=\columnwidth]{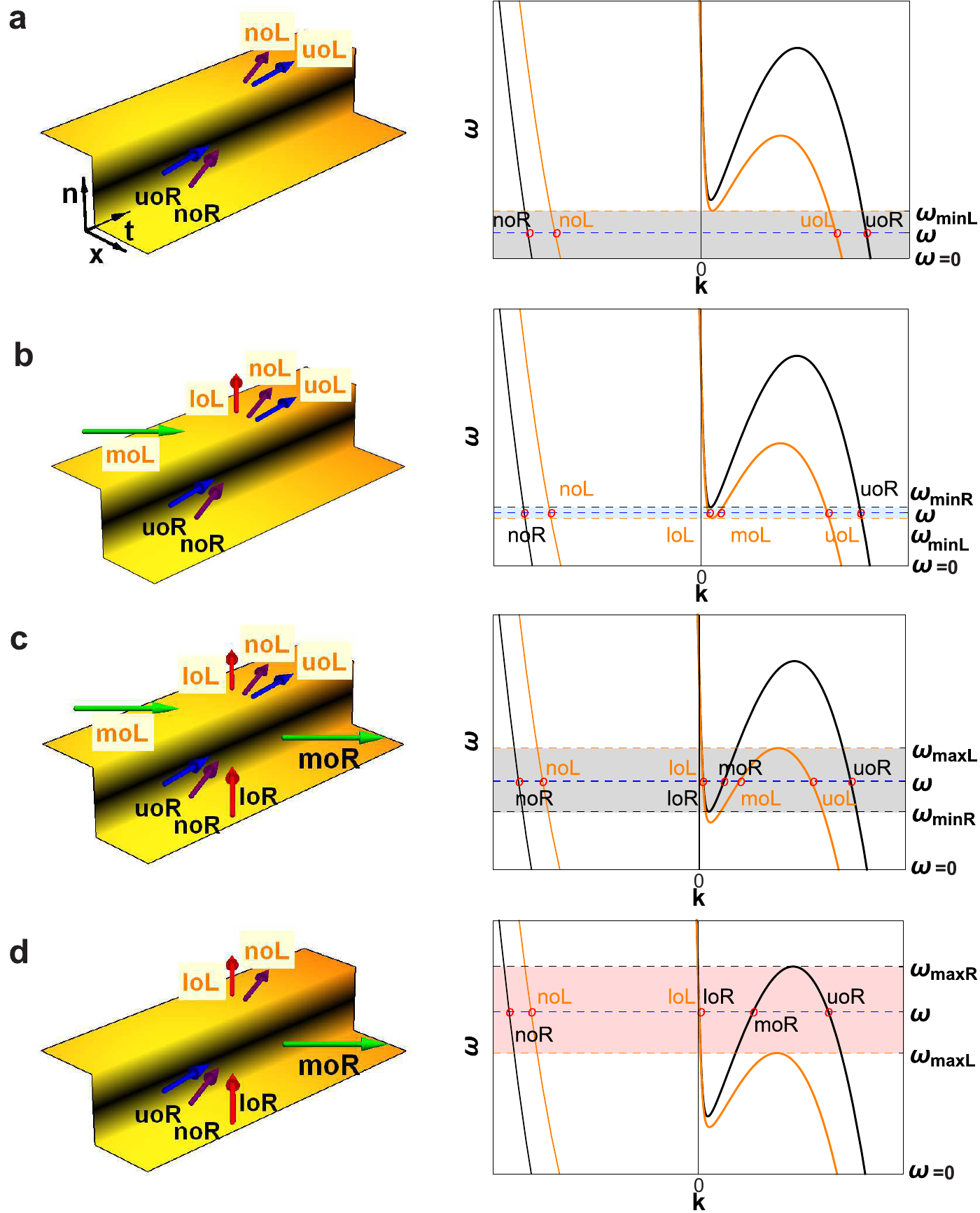}
\caption[Kinematic scenarios at the RIF]{Possible kinematic scenarios at the RIF.
Time (left) and frequency (right) illustrations of propagating modes are shown for different comoving frequencies $\omega$ (\textbf{a}-\textbf{d}).
The black (orange (light grey)) dispersion curve corresponds to the low (high) index region.
Single-frequency modes of $\omega$ are identified by intersections with the blue dashed line.
Possible kinematic scenarios, analogous to changes of spacetime curvature: \textbf{a}, horizonless scenario ($\omega<\omega_{minL}$); \textbf{b}, white hole scenario ($\omega_{minL}<\omega<\omega_{minR}$); \textbf{c}, horizonless scenario ($\omega_{minR}<\omega<\omega_{maxL}$); \textbf{d}, black hole scenario ($\omega_{maxL}<\omega<\omega_{maxR}$).\label{fig:kinematics}}
\end{figure}

The RIF separates two regions of homogeneous refractive index whose dispersion is modelled by (cf. App \ref{app:FT})
\begin{equation}
\label{eq:SellmDispRelMFbody}
c^2k^2=\omega^2+\sum_{i=1}^3\frac{4\pi\kappa_i\gamma^2\left(\omega+uk\right)^2}{1-\frac{\gamma^2\left(\omega+uk\right)^2}{\Omega_i^2}},
\end{equation}
with $\omega$ and $k$ the frequency and wavenumber in the frame co-moving with the RIF at speed $u$ ($\gamma=[1-u^2/c^2]^{-1/2}$).
$\Omega_i$ and $\kappa_i$ are the medium resonant frequencies and elastic constants.
The change in index, $\delta n$, at $x=0$ between the two regions, is modelled by a change in $\kappa_i$ and $\Omega_i$ in \eqref{eq:SellmDispRelMFbody}.
As illustrated in the dispersion diagrams of Fig.\ref{fig:kinematics}, the step manifests itself as a change in the dispersion relation between the low ($x>0$, black curve) and high ($x<0$, orange (light grey) curve) refractive index regions.

The two parts of the optical branch, shown in Fig.\ref{fig:kinematics} for frequencies of relevance, represent negative- (positive-) norm modes on the left (right) in the diagram.
Modes are marked with open circles.
For all $\omega$, there is one negative-norm mode (thin) and, in addition, either one or three positive-norm modes (thick).
We refer to frequency intervals of  $\omega$ with three positive norm modes as \textit{subluminal intervals} (SbLIs): $\left[\omega_{min L},\omega_{max L}\right]$ and $\left[\omega_{min R},\omega_{max R}\right]$. 
Inside a SbLI, one of the four mode solutions has a positive group velocity $\frac{\partial\omega}{\partial k}$ in the moving frame, moving rightwards.
We call this mode `mid-optical' --- `\textit{moL}' on the left, and `\textit{moR}' on the right of the boundary.
Note that in no other mode may light propagate with positive group velocity.
There is also the `low optical' mode \textit{loL}/\textit{R}, the `upper optical' mode \textit{uoL}/\textit{R}, and the `negative optical' mode \textit{noL}/\textit{R}. 
Beyond the SbLIs, \textit{i.e.} for $\omega\leq \omega_{minL/R}$ or $\omega\geq \omega_{maxL/R}$, only two propagating modes remain.
Thus, in growing order of $\omega$ we find the following distinct kinematic scenarios at the step:
\begin{enumerate}[-]
\item Fig.\ref{fig:kinematics} \textbf{a}. Two optical propagating modes (\textit{noL/R, uoL/R}) are present, with negative group velocities in the moving frame on either side. Light in all modes may only move to the left.
\item Fig.\ref{fig:kinematics} \textbf{b}. On the left of the interface, four optical propagating modes are present (\textit{noL, loL, moL, uoL}) whilst only two modes (\textit{noR, uoR}) are present on the right.
Mode \textit{moL} has positive group velocity, whereas all other modes have negative group velocity.
Light can propagate into the boundary from the left, but cannot proceed further to the right.
The interface acts as a white hole horizon to light.
\item Fig.\ref{fig:kinematics} \textbf{c}. Four propagating modes (\textit{noL/R, loL/R, moL/R, uoL/R}) exist on either side of the interface.
Modes \textit{moL} and \textit{moR} have positive group velocity on the left and on the right of the RIF, respectively.
The RIF is not a one-way door.
\item Fig.\ref{fig:kinematics} \textbf{d}. On the left of the interface, two propagating modes are present (\textit{noL}, \textit{uoL}), whilst four optical modes are present on the right (\textit{noR, loR, moR, uoR}).
Light on the right can move in either direction, but on the left of the interface both modes have negative group velocity.
There is a one-way door from the right to the left: light experiences a black-hole horizon at the RIF.
\end{enumerate}
For the largest $\omega$, the kinematics return to the type of Fig.\ref{fig:kinematics} \textbf{a}.

Only in scenarios \textbf{b} and \textbf{d} is a subluminal region paired with a superluminal region, creating a horizon.
This is in analogy to the superluminal space flow in the interior region of a black- or white hole and the subluminal flow outside \cite{jacquet_influence_2020}.
The analogy to gravity is best understood when thinking of black hole emission by the Hawking effect: a positive norm mode --- Hawking radiation --- allows for energy to propagate away from the hole in the subluminal region, while its negative norm partner falls inside the horizon.
The black hole kinematics as illustrated in Fig.\ref{fig:HRstep} are realised in Fig.\ref{fig:kinematics} \textbf{d}: mode \textit{moR} allows for light to escape the horizon to the right, like Hawking radiation, and mode \textit{noL} is the infalling partner.

This analysis of kinematic scenarios is interesting because it shows that optical analogues may allow us to access  features of gravity physics other than wave motion at the event horizon of black holes.
The simple geometry of a localised step allowed us to identify the above kinematic scenarios.
A continuous RIF would also exhibit one of these four scenarios at any one point $x$.

In all 4 kinematic scenarios, the motion of the RIF in the dispersive medium creates a non-adiabatic modification of the refractive index in time $T$, which corresponds to a time-dependance of the spacetime curvature.
Thus, modes (even in the vacuum state) will scatter at the RIF.
We use the S-matrix formalism to describe this mode coupling.

\section{Analytical calculation of spectral densities and correlations \label{sec:analyticsmethods}}
The aim of the field theory summarised in appendix \ref{app:FT} is to enable the calculation of all output properties in terms of the scattering matrix $S$.
In \cite{Jacquet_quantum_2015}, we calculated the output spectral density in all output modes.
Paired production in these modes, which is at the heart of the Hawking effect, is best characterised by the photon number variances and covariances.
Here, we derive these quantities for an optical system for the first time.

In appendix \ref{subsec:GMsSM} we construct two basis sets of global modes (GMs) of the system: the \textit{in} and \textit{out} bases.
We calculate the spontaneous emission in \textit{out} modes, \textit{i.e.}, the  \textit{in} modes are in the vacuum state $\ket{0^{in}}$.
Because the system is in a stationary state, the photon flux for an \textit{out} mode $\alpha$ is given by the integral over the frequency correlation \cite{loudon_2000}:
\begin{equation}
\begin{split}
\label{eq:fluxop}
\phi^\alpha(\omega)&=\left \langle 0^{\mathrm{in}}|\hat{\phi}^\alpha(\omega)|0^{\mathrm{in}} \right \rangle \\ &= \int\limits_0^\infty \frac{d\omega'}{2 \pi} \,\left \langle 0^{\mathrm{in}} |\hat{a}^{\mathrm{out}\, \alpha \dagger}({\omega}) \hat{a}^{\mathrm{out}\, \alpha}({\omega'})|0^{\mathrm{in}} \right \rangle.
\end{split}
\end{equation}
The flux $\phi(\omega)$ is the dimensionless number of photons per unit time $\delta\tau$ and unit bandwidth $\delta\omega$ at $\omega$ in the moving frame.
We now replace the \textit{out}-mode operators in the frequency correlations using the scattering transformation \eqref{eq:operatortransfo} by \textit{in}-operators, which act directly on the \textit{in} vacuum state.
Using the frequency correlation \eqref{eq:2ndmom1} calculated in appendix \ref{subsec:corrfunctionsSM}, we express the photon flux by the $S$-matrix
\begin{equation}
\label{eq:fluxbody}
\phi^\alpha(\omega)=\frac{1}{2 \pi}\sum_{\beta \notin \{\alpha\}}  {\left|S_{\alpha \beta}(\omega)\right|^2}.
\end{equation}
$\{\alpha\}$ is the set of modes which have a positive (negative) norm if $\alpha$ is of positive (negative) norm.
The photon flux results from the scattering of \textit{in} modes into \textit{out} modes of opposite sign of norm.
The number of emitted photons depends solely on the scattering matrix $S$, bandwidth and interaction time.

The total photon number is obtained by integrating over all frequencies and time.
To obtain the photon number $\hat{N}^\alpha_t$ over a limited `detected' frequency interval in this stationary mode conversion process, the field operators have to be constrained to this bandwidth. 
This is achieved via the spectral filter function $t(\omega)$, which takes the value of 1 if the frequency is in the detected interval and 0 otherwise.
Then
\begin{equation}
\label{eq:hatN}
\hat{N}^\alpha_t=\tau \bigg(\int\limits_0^\infty \!\! \frac{\mathrm{d}\omega}{\sqrt{2 \pi}} t(\omega) \hat{a}^{\mathrm{out}\,\alpha}(\omega)\bigg)^{\dagger} \bigg(\int\limits_0^\infty \!\! \frac{\mathrm{d}\omega}{\sqrt{2 \pi}} \\ t(\omega) \hat{a}^{\mathrm{out}\,\alpha}(\omega)\bigg).
\end{equation}
$\tau$ is the interaction time in the moving frame.
From \eqref{eq:hatN}, one may calculate other important quantities, such as \eqref{eq:fluxbody} and the photon-number correlations across the spectrum.

Correlations in photon number between modes $\alpha$ and $\alpha'$ are found by calculating the normally ordered covariance of the photon numbers $\hat{N}_1^\alpha$ and $\hat{N}_2^{\alpha'}$ on detectors 1 and 2:
\begin{equation}
\label{eq:covariance}
\begin{split}
 \mathrm{cov}(\!\hat{N}^{\alpha}_1\!,\hat{N}^{\alpha'}_2\!)\!&=\!\tau^2\!\!\! \iiiint\limits_0^\infty \! \frac{d\omega ... d\omega'''}{(2\pi)^2}t_1^{*}(\omega) t_2^{*}(\omega') t_2(\omega'') t_1(\omega''')\\
 &\!\!\!\!\!\!\!\! \big\langle\hat{a}^{\alpha \dagger}(\omega) \hat{a}^{\alpha' \dagger}(\omega') \hat{a}^{\alpha'}({\omega''}) \hat{a}^{\alpha}({\omega'''}) \big \rangle\,
\! -\! \big\langle \hat{N_1}^\alpha \big\rangle \big\langle \hat{N_2}^{\alpha'}\!\big\rangle.
\end{split}
\end{equation}
In \eqref{eq:covariance} the expectation value is again taken with respect to the \textit{in} vacuum state $\left|0^{\mathrm{in}}\right \rangle$ and $t_1$ and $t_2$ are the filters for detectors 1 and 2, respectively.
This defines the spectral intervals $\Delta_1$ and $\Delta_2$ over which these detectors collect photons.
We calculate the fourth order moment of the field $\left \langle \hat{a}^{\alpha \dagger}(\omega) \hat{a}^{{\alpha'} \dagger}({\omega'}) \hat{a}^{\alpha''}({\omega''}) \hat{a}^{\alpha'''}({\omega'''}) \right \rangle$ in Appendix \ref{subsec:corrfunctionsSM}. Inserting the result \eqref{eq:4thmom2} into \eqref{eq:covariance} we find
\begin{widetext}
\begin{equation} 
\begin{split}
 \label{eq:covofS1}
\mathrm{cov}&(\hat{N}^\alpha_1,\hat{N}^{\alpha'}_2) =\left(\frac{\tau}{2 \pi}\right)^2 \int\limits_0^\infty \int\limits_0^\infty  d\omega \, d\omega' \bigg[  \delta_{A} \sum_{\beta, \beta' \notin {\{\alpha\}} }  \mathcal{S}^*_{\alpha \beta'}(\omega) \mathcal{S}^*_{\alpha' \beta'}(\omega) \mathcal{S}_{\alpha' \beta}(\omega') \mathcal{S}_{\alpha \beta}(\omega')t_1^{}(\omega) t_2^{}(\omega) t_2(\omega') t_1(\omega') \\&+\delta_{S} \sum_{\beta, \beta' \notin \{{\alpha}\} }  \mathcal{S}^*_{\alpha \beta'}(\omega) \mathcal{S}^*_{\alpha' \beta}(\omega') \mathcal{S}_{\alpha' \beta'}(\omega) \mathcal{S}_{\alpha \beta}(\omega')t_1^{}(\omega) t_2^{}(\omega') t_2(\omega) t_1(\omega')  \\&+ \sum_{\beta\notin \{{\alpha}\}}\sum_{\beta'\notin \{{\alpha'}\}}  |S_{\alpha \beta}(\omega)|^2 |S_{\alpha' \beta'}(\omega')|^2 |t_1(\omega)|^2 |t_2(\omega')|^2 \bigg]-\big\langle \hat{N_1}^\alpha \big\rangle \big\langle \hat{N_2}^{\alpha'} \big\rangle .
\end{split}
\end{equation}
\end{widetext}

Here $\mathcal{S}_{\alpha \beta}$ is identical (complex conjugate) to $S_{\alpha \beta}$ if mode $\alpha$ is of positive (negative) norm.
$\delta_S, \, (\delta_A)$ equals unity if the correlated modes $\alpha$ and $\alpha'$ have identical (opposite) norm and zero otherwise.
Reverting from $\mathcal{S}$ back to $S$, we observe that in the $\delta_A$-term two $\mathcal{S}$ elements have to be conjugated and in the $\delta_S$-term either all or none. Since the result, as well as $t_i(\omega)$, are real, the first two sums in \eqref{eq:covofS1} are equal. Hence we observe that the mutually exclusive and exhaustive $\delta_S$ and $\delta_A$ let us combine the first two terms into one and the last two terms cancel due to \eqref{eq:operatortransfo}, \eqref{eq:fluxbody} and \eqref{eq:hatN}:
\begin{equation} \label{eq:covofS} \mathrm{cov}(\hat{N}^\alpha_1,\hat{N}^{\alpha'}_2) =\left(\frac{\tau}{2 \pi}\right)^2 \sum_{\beta \notin {\{\alpha\}} }\bigg| \int\limits_\Delta d\omega \,{S}^*_{\alpha \beta}(\omega) {S}_{\alpha' \beta}(\omega) \bigg| ^2.
\end{equation}
$\Delta$ is the spectral overlap (moving frame frequencies) of the two spectral intervals $\Delta_1$ and $\Delta_2$ of the two detectors. This result allows us to quantify the spectrally resolved photon number correlations of any stationary process in quantum optics. The correlations are contained in the scattering matrix and are dependent on the spectral overlap of the detectors regarding the investigated mode. 

We also calculate the variance $\mathrm{var}(\hat{N}^\alpha_1)= \langle \hat{N}^\alpha_1 \hat{N}^\alpha_1 \rangle - \langle \hat{N}^\alpha_1 \rangle^2  $.
Using the not normally ordered $4^{th}$ moment derived in \eqref{eq:nnovariance} of App. \ref{subsec:corrfunctionsSM} in a not normally ordered expression equivalent to \eqref{eq:covariance}, and with $\alpha=\alpha'$, we obtain
\begin{equation}
\label{eq:variance}
\mathrm{var}(\hat{N}^\alpha_1)=\langle \hat{N}^\alpha_1 \rangle\left(\langle \hat{N}^\alpha_1 \rangle+\frac{\tau \Delta_1}{2 \pi}\right).
\end{equation}
 
The photon-flux Pearson correlation coefficient between detectors 1 and 2, corresponding to modes $\alpha$ and $\alpha'$.
It is
\begin{widetext}
\begin{multline}
\label{eq:corrcoefficient}
\textrm{C}( \hat{N}^{\alpha}_1, \hat{N}^{\alpha'}_2)
\equiv\frac{\mathrm{cov}(\hat{N}^\alpha_1,\hat{N}^{\alpha'}_2)}{\left(\mathrm{var}(\hat{N}^\alpha_1)\,\mathrm{var}(\hat{N}_2^{\alpha'})\right)^{1/2}}\\
= \frac{  \bigg|\sum\limits_{\beta \notin {\{\alpha\}} } \int\limits_\Delta \!d\omega \,{S}^*_{\alpha \beta} {S}_{\alpha' \beta} \bigg| ^2}{\bigg[ \left( \sum\limits_{\beta \notin {\{\alpha\}} }\int\limits_{\Delta_1} \!\!d\omega |{S}_{\alpha \beta}|^2 \right)\left( \sum\limits_{\beta \notin {\{\alpha\}} }\int\limits_{\Delta_1} \!\!d\omega |{S}_{\alpha \beta}|^2+\Delta_1\right) \left( \sum\limits_{\beta \notin {\{\alpha'\}} }\int\limits_{\Delta_2} \!\!d\omega |{S}_{\alpha' \beta}|^2\right) \left( \sum\limits_{\beta \notin {\{\alpha'\}} }\int\limits_{\Delta_2} \!\!d\omega |{S}_{\alpha' \beta}|^2+\Delta_2\right) \bigg] ^{1/2}}\\
= \frac{\Delta^2}{\Delta_1 \Delta_2}\frac{  \bigg|\sum\limits_{\beta \notin {\{\alpha\}} }  \,{S}^*_{\alpha \beta} {S}_{\alpha' \beta} \bigg| ^2}{\bigg[ \sum\limits_{\beta \notin {\{\alpha\}} } |{S}_{\alpha \beta}|^2 \left( \sum\limits_{\beta \notin {\{\alpha\}} } |{S}_{\alpha \beta}|^2+1\right)  \sum\limits_{\beta \notin {\{\alpha'\}} } |{S}_{\alpha' \beta}|^2 \left( \sum\limits_{\beta \notin {\{\alpha'\}} } |{S}_{\alpha' \beta}|^2+1\right) \bigg] ^{1/2}}.
\end{multline}
\end{widetext}

In the last step we have assumed the scattering matrix to change little across a narrow detection bandwidth.
The normalisation ensures that $|\textrm{C}( \hat{N}^{\alpha}_1, \hat{N}^{\alpha'}_2)|\leq1$. 
Note that this also means that $\textrm{C}$ is independent of the photon flux.
The correlations generated by the scattering are entirely positive, indicating their origin of entangled photon pair generation from the vacuum. From \eqref{eq:corrcoefficient}, we can easily extract the (narrow bandwidth) self-correlation with matched filters
\begin{equation}
\label{eq:selfcorr}
\textrm{C}( \hat{N}^{\alpha}_1, \hat{N}^{\alpha}_1)
=\frac{\langle \hat{N}^{\alpha}_1 \rangle^2}{\mathrm{var}(\hat{N}^{\alpha}_1)}=1-\left(\frac{\mathrm{var}(\hat{N}^{\alpha}_1)}{\langle \hat{N}^{\alpha}_1 \rangle \frac{\tau \Delta}{2 \pi}}\right)^{-1},
\end{equation}
which is a measure of the noise $\mathrm{var}(\hat{N}^{\alpha}_1)$ in mode $\alpha$  relative to the Poisson noise variance of  $\langle \hat{N}^{\alpha}_1 \rangle \frac{\tau \Delta}{2 \pi}$. Using \eqref{eq:variance}, we see that $\textrm{C}$ increases from 0 (vacuum) and approaches 1 (maximum noise) for increasing photon numbers.

Furthermore, we calculate the second order correlation function $g^{(2)}_{\alpha \alpha'} = \frac{\langle:\hat{N}^{\alpha}_1 \hat{N}^{\alpha'}_2  : \rangle}{\langle \hat{N}^{\alpha}_1 \rangle\langle \hat{N}^{\alpha'}_2 \rangle}$.
Again assuming narrowband detection, we obtain:
\begin{equation}
\label{eq:gtwoaap}
	g^{(2)}_{\alpha \alpha'} = \frac{\Delta^2}{\Delta_1 \Delta_2} \left[1+ \frac{ |\sum\limits_{\beta \notin {\{\alpha\}} }  \,{S}^*_{\alpha \beta} {S}_{\alpha' \beta} |^2}{\sum\limits_{\beta \notin {\{\alpha\}} } |{S}_{\alpha \beta}|^2\sum\limits_{\beta \notin {\{\alpha'\}} } |{S}_{\alpha' \beta}|^2}\right]
\end{equation}
as well as the single detector correlation
\begin{equation}
\label{eq:gtwoaa}
	g^{(2)}_{\alpha \alpha} = 2.
\end{equation}
Here we recover the expected result that the scattering induces correlated noise between the two detectors, both of which detect the statistics of chaotic light.
These relations describe output observables in any process described by the scattering matrix but are particularly tailored to analogue gravity in dispersive systems, in which context they appear for the first time.

The photon numbers and correlations up to this point are calculated in the \textit{moving frame} where the system is stationary.
However, in an experiment, measurements are performed in the \textit{laboratory frame}, so we now proceed to derive the flux there.
Photon numbers and time-bandwidth products $\tau \Delta$ are frame invariant.
Therefore, re-interpreting $\tau$ and $\Delta$ as the \textit{laboratory} interaction time and detection bandwidth, the \textit{laboratory} frame correlation coefficient of modes $\alpha$  and $\alpha'$ is identical to Eq.\eqref{eq:corrcoefficient}.
All other relations containing photon number operators can be interpreted for the laboratory frame as well.

For light waves to reach the detector, the \textit{out} global modes (GMs) must have a \textit{positive} laboratory-frame group velocity. 
As can be seen in Fig.\ref{fig:labdisprel}, modes \textit{no} and \textit{uo} have positive group velocities.
So too does mode \textit{lo}, except if $K$ is negative (at very large $\omega$).
These modes are the \textit{out} GMs on the left. 
In the low-index region on the right, the only \textit{out} GM, \textit{i.e.}, propagating \textit{away} from the RIF and ahead of it, is mode \textit{moR}.
Thus, when calculating the laboratory frame spectral density at optical frequencies, we expect to detect contributions from the \textit{out} GMs \textit{loL} (at low $\omega$), \textit{moR}, \textit{uoL} and \textit{noL}.

 \begin{figure}[t]
     \centering
     \includegraphics[width=.85\columnwidth]{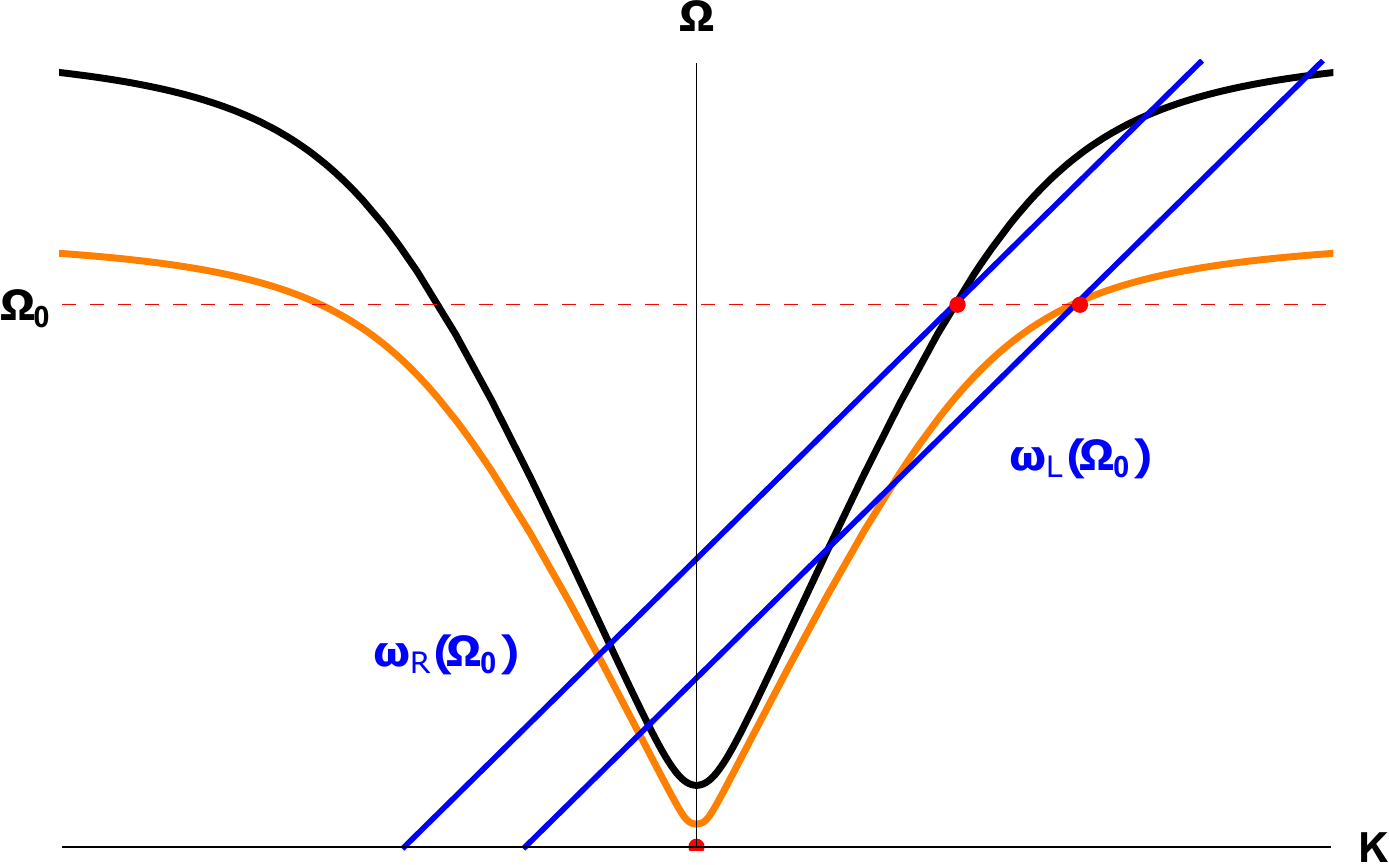}
     \caption[Mode contributions in the laboratory frame]{Diagram of possible mode contributions in the laboratory frame: the optical frequency branch is shown in the laboratory frame on the left (high refractive index --- orange (light grey)) and the right (low refractive index --- black) of the RIF (see Fig.\ref{fig:HRstep}). For a given laboratory frame frequency, $\Omega_0$, modes corresponding to solutions of \eqref{eq:SellmDispRelMF} at two moving frame frequencies $\omega_R(\Omega_0)$ and $\omega_L(\Omega_0)$ may contribute to the emission. These local modes define \textit{out} GMs and their contributions add up to the laboratory frame spectral density \eqref{eq:LSDlambda} and photon-number correlation \eqref{eq:corrcoefficient} at $\Omega_0$.\label{fig:lsdalgo}}
 \end{figure}

The \textit{moving frame} flux density $\phi^{\alpha}({\omega})$ in \textit{out} GM $\alpha$ is \eqref{eq:fluxbody}. 
The \textit{laboratory frame} flux density $\Phi^{\alpha}(\Omega)$ is obtained from this by \cite{finazzi_quantum_2013}
\begin{equation}
\label{eq:LFrate}
\Phi^{\alpha}({\Omega})=\left(1-\frac{u}{v_{g}(\Omega)}\right)\phi^{\alpha}({\omega}).
\end{equation}
The total spectral density at $\Omega$, $\Phi(\Omega)$, is found by adding contributions of all GMs:
\begin{equation}
\label{eq:LSD}
\Phi({\Omega})=\sum_\alpha \Phi^{\alpha}({\Omega}).
\end{equation}

Here, we have to consider that a laboratory-frame detector might detect more than a single mode $\alpha$.
Assuming the detector is sensitive to a frequency $\Omega_0$, then we can read off the laboratory frame dispersion diagram (Fig.\ref{fig:lsdalgo}) that there are two modes with $\Omega_0$ and opposite $K$. 
Note that there is another pair of solutions at $-\Omega_0$.
Out of the four solutions exactly one has positive group velocity and positive moving frame frequency $\omega$.
Thus the detector frequency corresponds to a unique mode in general, although we remark that two exceptions are possible: (a) The detected frequency interval could contain the boundary frequency separating two modes, \textit{e.g.} \textit{moR} and \textit{uoL}.
In this case the interval is reduced to the detected mode with group velocity away from the RIF.
(b) Two out-GMs from either side of the RIF might share a laboratory frequency $\Omega_0$  (Fig.\ref{fig:lsdalgo}).
In this case they typically do not share the same moving frame frequency ($\omega_{R}(\Omega_0)\neq\omega_{L}(\Omega_0)$) and the covariances and variances of the modes add up, as for incoherent fields.

The wavelength equivalent to $\Phi(\Omega)$ \eqref{eq:LSD} is
\begin{equation}
\label{eq:LSDlambda}
\Phi_\lambda({\lambda})= \frac{2\pi c}{\lambda^2} \sum_\alpha \left(1-\frac{u}{v_g\left(\frac{2\pi c}{\lambda}\right)}\right)\phi^{\alpha}\left(\omega\right),
\end{equation}
with $\omega=\gamma\left(\frac{2\pi c}{\lambda}-u K^\alpha\right)$; $K^\alpha$ the laboratory frame wavenumber; $\alpha$ the modes that contribute to the emission at $\lambda$, and $\phi^{\alpha}\left(\omega\right)$ their moving frame flux density \eqref{eq:fluxbody}; $v_g$ the mode's group velocity in the laboratory frame.

Equation \eqref{eq:LSDlambda} completes this section, where we formulated the important output observables for optical analogue gravity in both frames.
These quantities are the tools we will use in section \ref{sec:CS} to characterise and typify the emitted field in the moving and laboratory frames.

\section{Analytical calculation of the scattering matrix\label{sec:analyticsSmethods}}

As we have seen in the previous section, the quantum field for all modes explicitly follows from the scattering matrix, and we now build on the field theory of appendix \ref{app:FT} to derive the latter analytically for the first time.
We calculate the scattering matrix at all frequencies (from all kinematic scenarios) and include evanescent waves in the calculation, and thus our method lends itself to considerations of more general RIF profiles.

We continue to use the step in the index, as in Fig.\ref{fig:HRstep}.
For a monochromatic field of frequency $\omega$, the canonical conjugate momenta $\Pi_A$ and $\Pi_{P_i}$ as well as their first spatial derivatives can be expressed by the electromagnetic potential $A$, the polarisation fields $P_i$ as well as their derivatives in a homogeneous region using \eqref{eq:fieldmotioneqMFFT1}, \eqref{eq:fieldmotioneqMFFT2} and \eqref{eq:fieldmotioneqMFFT4} (App. \ref{app:FT}).
Therefore the field vector $\vec{V}=(A\ P_1\ P_2\ P_3\ \Pi_A\ \Pi_{P_1}\ \Pi_{P_2}\ \Pi_{P_3})^T$ can be reexpressed by the vector $\vec{W}= (A, P_1, P_2, P_3, A', P'_1, P'_2, P'_3)^T$.
The prime denotes the spatial derivative of the field.
$\vec{W}$ will allow us to simplify the matching conditions below.

$\vec{V}$ and $\vec{W}$ are related by Eqs.\eqref{eq:fieldmotioneqMFFT1}-\eqref{eq:fieldmotioneqMFFT4} as:
\begin{widetext}
\begin{equation}
\label{eq:Umatrix}
\vec{V}=
\left(\begin{array}{cccccccc}
1&0&0&0&0&0&0&0 \\
0&1&0&0&0&0&0&0 \\
0&0&1&0&0&0&0&0 \\
0&0&0&1&0&0&0&0 \\
i\frac{\omega}{4\pi c^2}&0&0&0&0&0&0&0\\
\frac{\gamma}{c}&-i\frac{\omega\gamma}{\kappa_1\Omega_1^2}&0&0&0&-\frac{u\gamma^2}{\kappa_1\Omega_1^2}&0&0\\
\frac{\gamma}{c}&0&-i\frac{\omega\gamma}{\kappa_2\Omega_2^2}&0&0&0&-\frac{u\gamma^2}{\kappa_2\Omega_2^2}&0\\
\frac{\gamma}{c}&0&0&-i\frac{\omega\gamma}{\kappa_3\Omega_3^2}&0&0&0&-\frac{u\gamma^2}{\kappa_3\Omega_3^2}\\
\end{array}\right)
\vec{W},
\end{equation}
\end{widetext}
for a field at frequency $\omega$.
We call the matrix in \eqref{eq:Umatrix} $\mathcal{U}$, and note that $Det(\mathcal{U})=0$.
We make \eqref{eq:Umatrix} applicable for global modes by defining $\mathcal{U}=\mathcal{U}_L\,\theta(-x)+\mathcal{U}_R\,\theta(x)$.
We may combine eight plane wave modes $\alpha$ ($\alpha\in\{\alpha_1,...,\alpha_8\}$) into a mode set as a matrix $W$
\begin{equation}
\label{eq:wmat}
W_{L/R}=\left(\vec{W}_{L/R}^{\alpha_1}\ \vec{W}_{L/R}^{\alpha_2}\ ... \ \vec{W}_{L/R}^{\alpha_8}\right),
\end{equation}
with $\vec{W}^\alpha$'s ordered by laboratory-frame frequency.
L/R corresponds to a basis set of local modes on either side of the boundary at $x=0$.
This is similar to \eqref{eq:fieldtransfo}.

\begin{figure*}[ht]
    \centering
    \includegraphics[width=.85\linewidth]{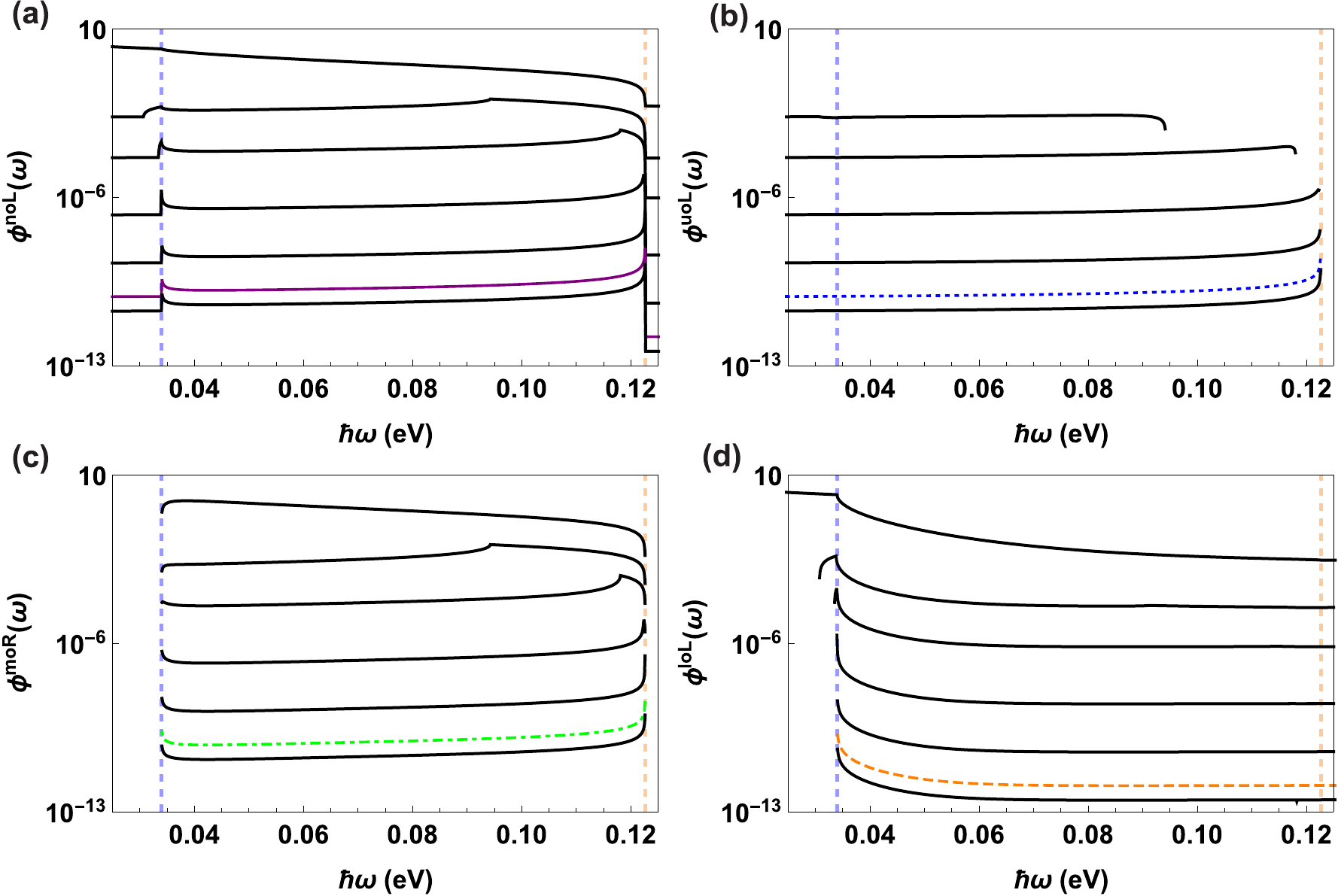}
    \caption[Emission spectra of the four modes in the moving frame]{Photon flux \eqref{eq:fluxbody} of each optical mode in the moving frame for varying step heights (from bottom to top, $\delta n=10^{-6},\,2\times 10^{-6}$ (coloured lines), $10^{-5},\, 10^{-4},\, 10^{-3},\, 10^{-2},\, 10^{-1}$): \textbf{(a)} \textit{noL}, \textbf{(b)}  \textit{uoL}, \textbf{(c)} \textit{moR}, \textbf{(d)} \textit{loL} (cf. Fig.\ref{fig:labdisprel}). The vertical blue- and orange-dashed lines identify the limits of the WHI and BHI, respectively, for $\delta n=2\times 10^{-6}$. \label{fig:mfflux}}
\end{figure*}

In order to describe the mode coupling at the RIF with the scattering matrix formalism, we construct the global modes (GMs) of the inhomogeneous system, introduced in section \ref{subsec:GMsSM} of App. \ref{app:FT}.
These GMs are linear combinations of the plane wave solutions $\vec{W}^\alpha$, the local modes (LMs), for each homogeneous medium on either side of the RIF.
There are complete sets of global \textit{in} modes,
\begin{equation}
\label{eq:GMsets1}
W^\mathrm{in}(x)=W_L(x)\, \sigma_L^\mathrm{in} \,\theta(-x)+W_R(x)\, \sigma_R^\mathrm{in} \,\theta(x),
\end{equation}
and global \textit{out} modes,
\begin{equation}
\label{eq:GMsets2}
W^\mathrm{out}(x)=W_L(x)\, \sigma_L^\mathrm{out} \,\theta(-x)+W_R(x)\, \sigma_R^\mathrm{out} \,\theta(x).
\end{equation}
Each global \textit{in} (\textit{out}) mode contains only one LM, whose energy flux is directed into (out of) the boundary.
Because there are $16$ LMs connected at the index boundary, there exist 16 GMs: 8 \textit{in} and 8 \textit{out}.
In \eqref{eq:GMsets1} and \eqref{eq:GMsets2}, each $8\times8$ $\sigma$-matrix contains the coefficients of the 8 local modes for the 8 global modes.

The mode decompositions of GMs into LMs on the left and on the right of the interface, respectively, are related by the matching conditions at the interface.
For \textit{in} and \textit{out} modes at $x=0$, we write
\begin{align}
\label{eq:matchINLeftRightSigma}
W^\mathrm{in}&=W_{L}\ \sigma_{L}^{in}=W_{R}\ \sigma_{R}^{in} \\
\label{eq:matchOUTLeftRightSigma}
W^\mathrm{out}&=W_{L}\ \sigma_{L}^{out}=W_{R}\ \sigma_{R}^{out}.
\end{align}

This leaves us with 64 unknowns in either \eqref{eq:matchINLeftRightSigma} or \eqref{eq:matchOUTLeftRightSigma}.
The $W$'s are known, so we can obtain all $\sigma$'s (we calculate an example in Appendix \ref{subsec:smfrommcsSM}). 
We calculate the scattering matrix by relating the \textit{in} and \textit{out} GMs as:
\begin{equation}
\label{eq:Wtransfo}
W^\mathrm{in}(x)=W^\mathrm{out}(x) \,S.
\end{equation}
Insertion of \eqref{eq:GMsets1} and \eqref{eq:GMsets2} into \eqref{eq:Wtransfo} yields $S$:
\begin{equation}
\label{eq:smatrixsigma}
S={\sigma^{out}_{L}}^{-1}\sigma^{in}_{L}={\sigma^{out}_{R}}^{-1}\sigma^{in}_{R}.
\end{equation}

Now we have demonstrated how to completely characterize the scattering off a RIF analytically by calculation of the scattering matrix.
In Appendix \ref{subsec:smfrommcsSM}, we exemplify one mode configuration in detail (further calculations can be found in \cite{Jacquet_book_2018}), and explicitly derive the \textit{in} and \textit{out} $\sigma$ matrices, and the corresponding scattering matrix. In Appendix \ref{subsec:quasiunitaritySM}, we show that the scattering matrix is quasi-unitary, and thus correctly normalised.

\section{Spectra of spontaneous emission\label{sec:CS}}
We now put the analytical formulas developed in sections \ref{sec:anagrav}-\ref{sec:analyticsSmethods} into action to exemplify their use, and more importantly to study our dispersive, multimode system. 
The system is reduced to the essentials of analogue horizon physics, compared to more comprehensive and realistic systems.

We calculate the scattering matrix and compute 1- and 2-photon spectra of spontaneous emission.
Optical analogue experiments are different from their fluid-based counterparts (such as BECs \cite{munoz_de_nova_observation_2019}, water waves \cite{rousseaux_observation_2008,weinfurtner_measurement_2011,Rousseaux_PRL_2016} or fluids of light \cite{nguyen_acoustic_2015,Fontaine_disprel_2018,michel_superfluid_2018,jacquet_fluids_2020}) in that the reference frames are exchanged: the rest frame of the optical experiment corresponds to the frame of the moving fluid, and vice versa.
In both analogues, the measurements are performed in the laboratory frame.
Here, we present spectra and mode-correlation maps in both frames to predict observations in optics and to allow comparison to fluid-based analogues.

We continue with the example of light in bulk fused silica at a step-like RIF as in section \ref{sec:anagrav}.
The Sellmeier coefficients in \eqref{eq:SellmDispRelMFbody} are: $\kappa_{1,2,3}= 0.07142$, $0.03246$, and $0.05540$ for the elastic constants, and $\Omega_{1,2,3}=190.341 \, \mathrm{THz}$, $16.2047\, \mathrm{PHz}$, and $27.537\, \mathrm{PHz}$ for the resonance frequencies \cite{Agrawal_2012}.
Here, we present computations for various step heights, from $\delta n=10^{-6}$ to $\delta n=10^{-1}$.
We first consider a RIF moving at velocity $u=2/3c$ in bulk fused silica and later vary the velocity of the RIF (in section \ref{subsec:commentresults}).

\begin{figure*}[t]
    \centering
    \includegraphics{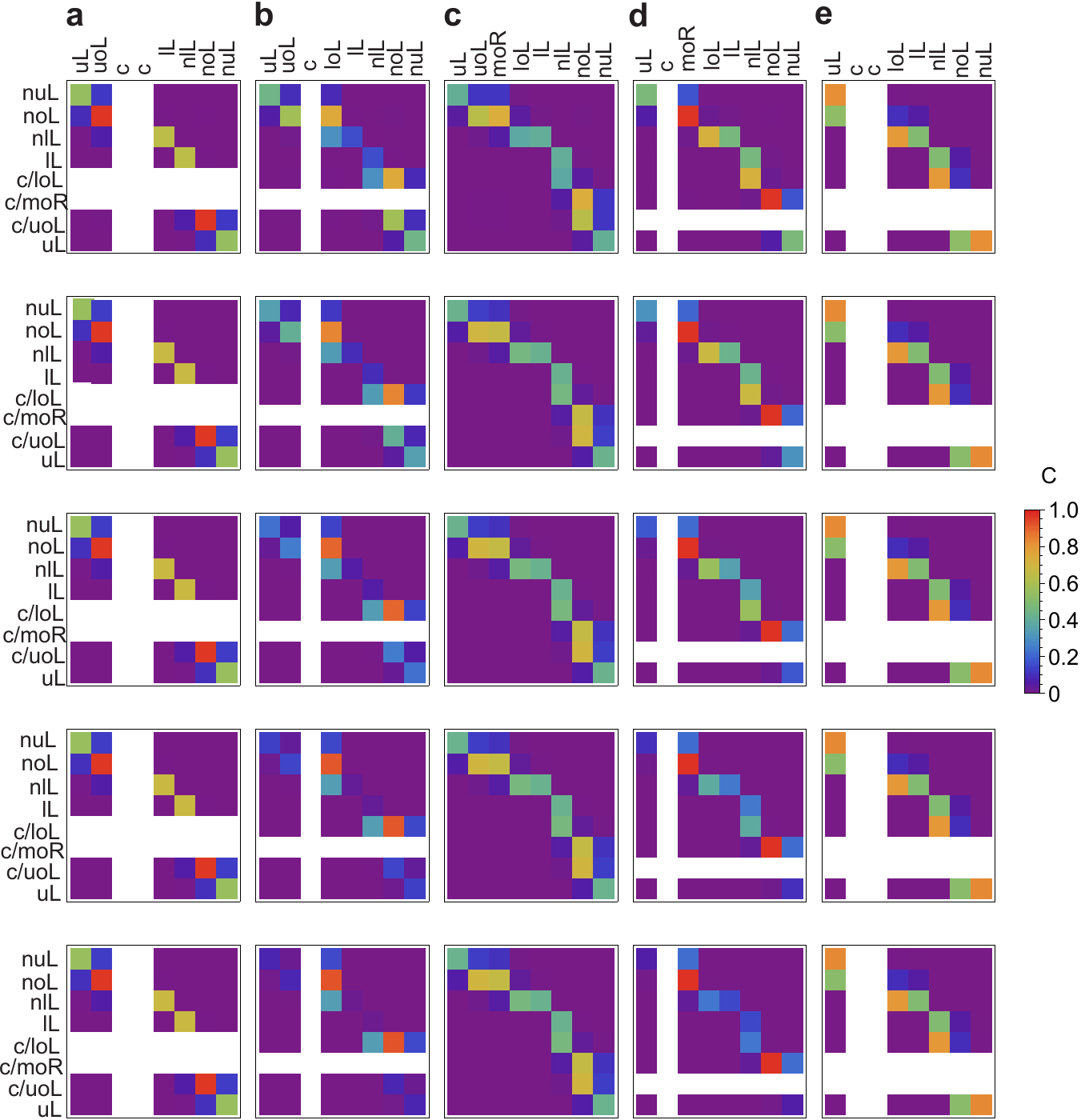}
    \caption[Mode-solutions cross correlations]{Matrices of photon number correlations \eqref{eq:corrcoefficient} between the eight outgoing modes for five typifying frequencies: columns \textbf{a}-\textbf{d} as in Fig.\ref{fig:kinematics} and \textbf{e} for high $\omega$, and rows for values of $\delta n$ (from bottom to top, $10^{-6},\,10^{-5},\, 10^{-4},\, 10^{-3},\, 10^{-2}$); $c$ --- a complex, unphysical mode; $uL,\, lL,\, nlL,\, nuL$ --- non-optical modes (see Fig.\ref{fig:labdisprel}).
\label{fig:mfcorr}}
\end{figure*}´

\subsection{Moving frame spectra\label{subsec:quantemcst}}
In the moving frame, the contributions of the different modes under the various kinematic scenarios are most apparent.
We use the formulas developed in sections \ref{sec:anagrav}-\ref{sec:analyticsSmethods} to calculate 1- and 2-photon spectra.
We also investigate the influence of the step height, which we change over a (realistic) range of 5 orders of magnitude, on the emission.

We start by computing the spontaneous photon flux \eqref{eq:fluxbody} in the moving frame in the four optical modes, see Fig.\ref{fig:mfflux} (\textbf{(a)} \textit{noL}, \textbf{(b)} \textit{uoL}, \textbf{(c)} \textit{moR} and \textbf{(d)} \textit{loL}) for various step heights $\delta n$.
The emission in all modes increases with step height.
The frequency intervals with horizons are located at the blue and orange dashed lines for the white hole interval (WHI) and black hole interval (BHI), respectively (kinematic scenario of Fig.\ref{fig:kinematics} \textbf{b} and \textbf{d}).
Except for the three highest step heights, the frequency intervals with event horizons are very narrow (\textit{e.g.} 1 and 10 $\mathrm{\mu}$eV for the WHI and BHI, respectively, at $\delta n=2\times10^{-6}$).
Here we observe the following emission peaks: modes \textit{loL} and \textit{noL} both form a peak at the WHI, whereas modes \textit{moR} and \textit{noL} both form a peak at the BHI.
Otherwise, modes \textit{moR} and \textit{uoL} sharply increase towards the WHI and BHI, respectively, but neither of them exists there.
Outside the horizon intervals the emission is spectrally wide and smooth.
We observe that horizons increase the emission.

Spectra in the horizon intervals assume the shape of a shark fin, best seen at higher step heights ($\delta n=10^{-3}$, $10^{-2}$) as the horizon intervals widen for emission into mode \textit{noL} at the WHI and BHI \textbf{(a)}, mode \textit{moR} at the BHI \textbf{(c)} and mode \textit{loL} at the WHI \textbf{(d)}.
This shape is created because on one side (large and small $\omega$) the emission cuts off by orders of magnitude as an emitting mode (\textit{moR} and \textit{loL}) ceases to exist outside the interval.
On the other side, the horizon condition cuts off because we approach the kinematic scenario of Fig.\ref{fig:kinematics} \textbf{c}, leading to an abrupt decrease in emission.
At the highest step height the BHI expands and the WHI disappears, leading to a more uniform emission.
The shape of the spectrum is formed by the change in kinematic scenario \cite{Jacquet_quantum_2015}.

Spontaneous emission occurs in pairs of photons, so we expect the photon number in any mode to be correlated with that in (at least) one other mode.
Partnered emission can be characterised further by computing the matrix of mutual photon number correlations between all modes.
Fig.\ref{fig:mfcorr} shows the photon-flux correlation coefficient \eqref{eq:corrcoefficient} of two modes at typifying frequencies $\omega$ and for $\Delta^2=\Delta_1\Delta_2$.
Correlations (\textit{e.g.} top row) are generally strongest between the optical modes.
Because \textit{noL} is the unique negative-norm optical mode, we find the strongest correlation between this mode and positive-norm optical modes.
We observe that the correlations are different if horizons exist (Fig.\ref{fig:mfcorr} columns \textbf{b}, \textbf{d}).
Over the WHI (BHI), there is a single large correlation between \textit{noL} and \textit{loL} (\textit{noL} and \textit{moR}), with other correlations being small or zero.
In both cases, the pairs of modes correspond to the Hawking radiation and the partner.
Importantly, partnered emission does not occur solely in the frequency intervals where the kinematics are analogous to the curvature of spacetime at a black- or white-hole horizon, but occurs in the other kinematic scenarios as well.
Without horizons (Fig.\ref{fig:mfcorr} \textbf{a}, \textbf{c}, \textbf{e}), significant correlations exist between typically three mode pairs simultaneously, indicating multipartite entanglement.
These involve non-optical modes, although the flux is weak \cite{Note1}.
As can be seen by comparing the columns of Fig.\ref{fig:mfcorr}, the magnitude of the correlation coefficients as well as their structure are surprisingly independent from the step height, although this varies over 5 orders of magnitude.

To summarise,  the flux of spontaneous emission is dominated by white- or black hole-horizon physics and drops significantly beyond that.
The spectral characteristic is a `shark fin' shape.
Over the analogue white- and black hole intervals,  paired two-mode emission at optical frequencies dominates.
These observations are robust over a wide range of step heights $\delta n$.

\subsection{Laboratory frame spectra\label{subsec:lfcalc}}
Using our formulas for observables further, we now compute laboratory frame spectra and identify the signature effects as observed in the laboratory frame.
Because we have found the signatures of horizon physics to be robust against changes in the step height $\delta n$, we now limit our discussion to $\delta n=2\times 10^{-6}$.

\begin{figure}[t]
    \centering
    \includegraphics[width=.95\columnwidth]{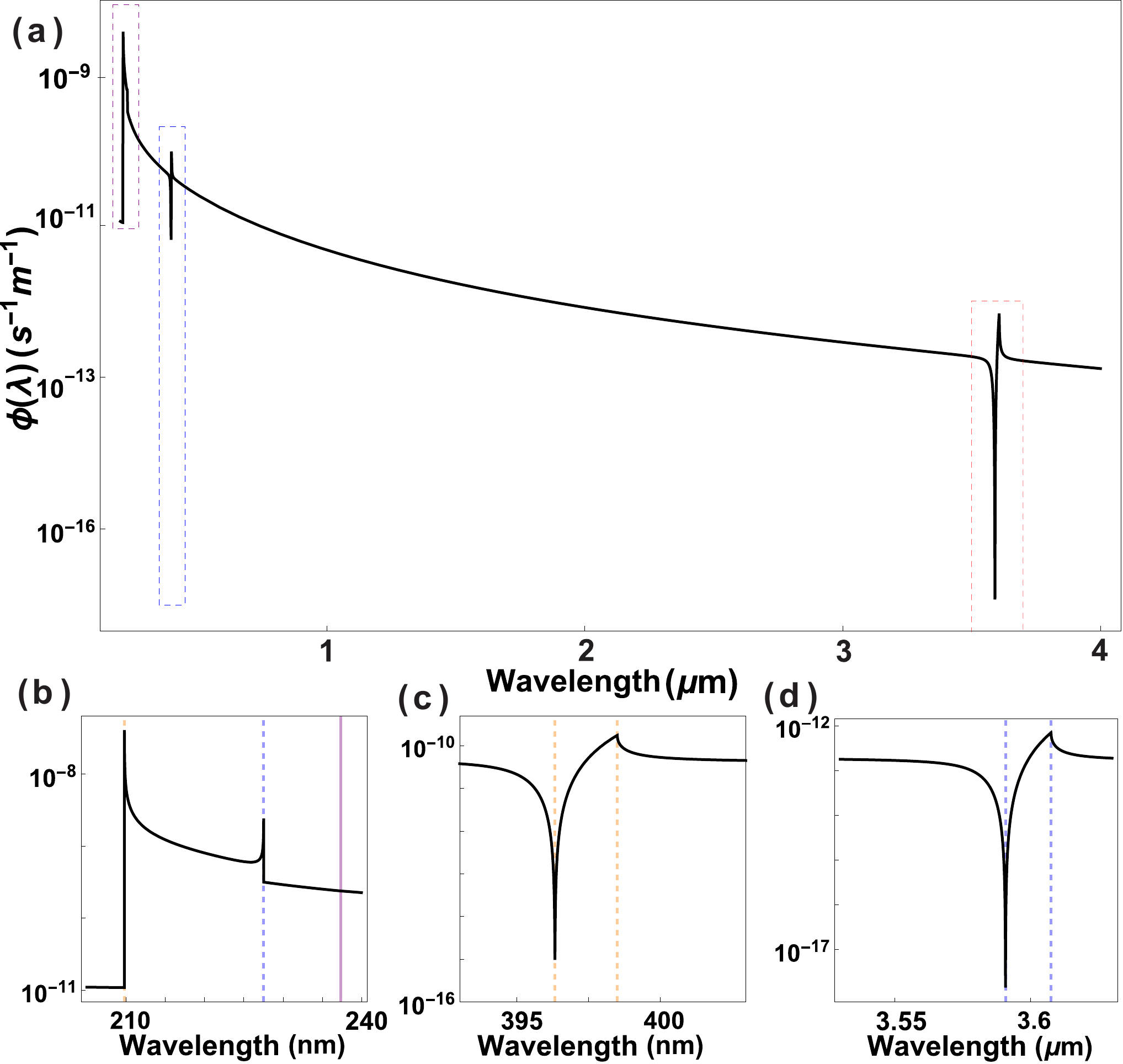}
\caption[Spectral density of  emission in the laboratory frame]{Spectral density of  emission in the laboratory frame \eqref{eq:LSDlambda}.
Emission from horizons of negative/positive norm: black hole $210\,$nm/$398\,$nm (dashed orange in \textbf{(b)} and \textbf{(c)}) and white hole $227\,$nm/$3.6\,\mu$m (dashed blue in \textbf{(b)} and \textbf{(d)}). There is a phase velocity horizon at $237$ nm (solid purple line in \textbf{(b)}).\label{fig:LSD}}
\end{figure}

The laboratory frame spectrum \eqref{eq:LSDlambda} is computed in Fig.\ref{fig:LSD}. 
Similar to its moving frame counterpart, it consists of intervals of white hole, black hole and horizonless emission.
Emission is sharply peaked over analogue white- and black-hole intervals, and dips down at wavelengths which are velocity-matched with the RIF.
These features are therefore ruled by the kinematic scenarios explored in Fig.\ref{fig:kinematics}.
The peak at $210$ ($227$) nm corresponds to black- (white-) hole emission into the optical negative-norm mode (\textit{noL}).
From there, the spectral density decreases towards $400$ nm.
At $396.34\,$nm, mode \textit{uoL} is velocity-matched to the RIF, becomes complex, and the emission drops.
The drop is limited to 5 orders of magnitude because at $396.33\,$nm black hole emission into mode \textit{moR} cuts on.
Finally, we observe peaks at $398\,$nm and around $3.6\,\mu$m that account for the black hole and white hole emission into positive norm modes \textit{moR} and \textit{loL}, respectively.
Note that, whereas in the moving frame the frequency sets the kinematic scenario, in the laboratory frame the spectral density may emerge from different scenarios.
All peaks in Fig.\ref{fig:LSD} \textbf{(b)}, \textbf{(c)} and \textbf{(d)} exhibit the signature `shark fin' structure of horizon emission.
Clearly, these are signature effects of horizon physics in dispersive (optical) media.

The various spectral peaks and dips we have identified have narrow linewidths: about $1\,$nm below $250\,$nm, $2\,$nm around  $400\,$nm and $17\,$nm at $3.6\,\mu$m.
They are expected to have strong spectral correlations.

\begin{figure*}[t]
\centering
\includegraphics[width=.65\linewidth]{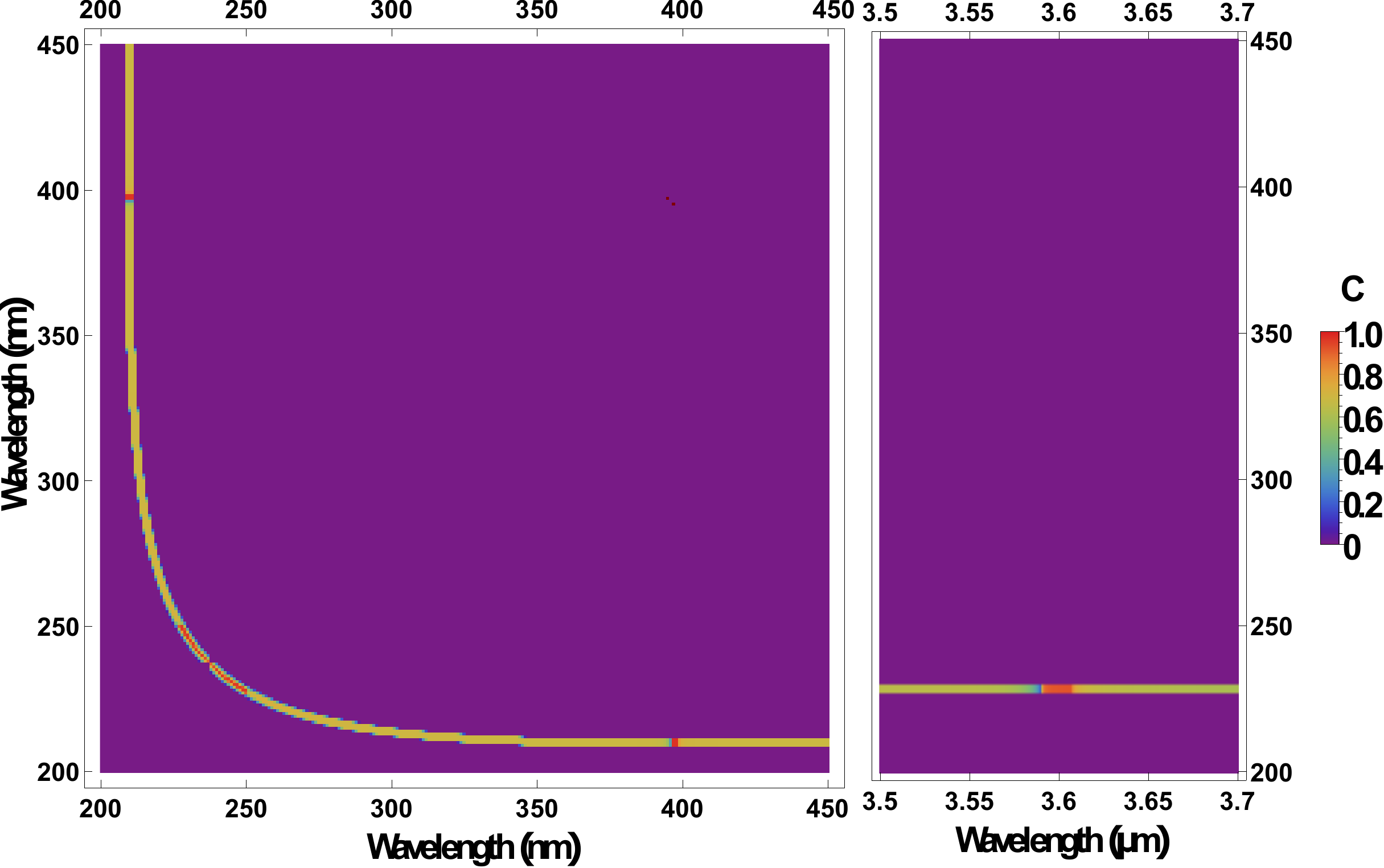}
\caption[Correlation map]{Photon number correlations in the laboratory: The RIF leads to broadband entangled pair production with a  near-unity coefficient at the BHI around ($397$ nm, $210$ nm) and the WHI around ($3.6\,\mu$m, $227$ nm).
\label{fig:lfcorr}}
\end{figure*}

Fig.\ref{fig:lfcorr} presents the photon number-correlation coefficient \eqref{eq:corrcoefficient} across the spectrum of laboratory frame emission.
Measuring these photon-number correlations convincingly reveals their vacuum fluctuations origin. 
Coefficients between different modes display a single continuous contour of significant photon number correlations across the entire spectrum. 
The contour indicates correlations between the negative-norm mode \textit{noL} (below $237\,$nm) and the positive-norm modes \textit{uoL} (between $237\,$nm and $396\,$nm) or \textit{moR} (beyond $396\,$nm).
The contour thus indicates where these mode pairs share a common moving frame frequency $\omega$. Note that this depends on dispersion in a nontrivial way and is not a hyperbolic relation. 
The correlation coefficients of Fig.\ref{fig:mfcorr} and Fig.\ref{fig:lfcorr} are mostly identical for identical frequencies $\omega$.
Along the contour in Fig.\ref{fig:lfcorr}, starting at $237$ mn, the  correlation coefficients gradually decrease as the wavelength of \textit{noL} decreases and the wavelength of mode \textit{uoL}, and subsequently of mode \textit{moR}, increases, up to a point around ($\lambda_{1},\,\lambda_2$)$=$($400\,$nm, $210\,$nm) where the black hole horizon is formed and the correlation coefficient suddenly peaks.
We observe the strongest coefficient, $0.97$,  at  ($397\,$nm, $210\,$nm) --- between modes \textit{moR} and \textit{noL}.
Partnered emission also dominates over the WHI ($3.6\,\mu$m, $227$ nm) with $C=0.92$ --- between modes \textit{loL} and \textit{noL}.
In both cases, the coefficients mainly deviate from unity because of weak correlations with non-optical modes ($nuL$ and $nlL$).
The quantum state is almost a pure two-mode squeezed vacuum \cite{jacquet_influence_2020}.
The strong correlation between outgoing modes only if separated by a horizon is characteristic of the Hawking effect.
On the diagonal, the self-coefficients \eqref{eq:selfcorr} characterize the photon number noise relative to Poisson noise.
These elements are very small as individual modes carry chaotic noise and the photon numbers are small.

\subsection{Robustness of the spectra against changes in dispersion\label{subsec:commentresults}}
Finally, we discuss the experimental implementation of this optical gravity analogue via the prism of the detection of Hawking radiation (HR).
The RIF might be generated by an optical pulse via the optical Kerr effect.
The velocity $u$ then corresponds to the group velocity of that pulse.
Changing $u$ by way of dispersion, \textit{e.g.} by shifting the pulse frequency, will therefore alter the wavelengths of the black- and white hole emission: these are found close to wavelengths which have the same group velocity as the RIF.

In Table \ref{tab:tableu} we show the laboratory-frame properties (wavelength, spectral density and spectral correlations) of white-hole and black-hole emission for four pulse group velocities (or wavelength $\lambda_c$): $u/c=2/3$ ($400$ nm), $2.04/3$ ($800$ nm), $2.05/3$ ($1260$ nm) and $2.04/3$ ($1990$ nm).
While $1260$ nm is the zero dispersion wavelength, where the group index assumes a maximum, $800$ nm and $1990$ nm share the same group velocity.
The table shows that overall emission wavelengths for the optical modes are shifted, but the correlation strengths are largely robust.
The flux strength changes moderately with the pulse velocity.
However, because the black hole emission into mode \textit{moR} is very close to the pulse wavelength, it will be difficult to distinguish it from photons of the pulse. Also, the white hole emission at infrared wavelengths is hard to detect.

When the group velocity of the RIF matches the group velocity of waves in the visible, the Hawking effect may create pairs of photons in the UV (the partner) and visible (HR).
This can be achieved by using a similar medium to ours, for example metamaterial waveguides, such as photonic crystal fibres \cite{philbin_fiber-optical_2008,choudhary_efficient_2012,Jacquet_book_2018,mclenaghan_compression_2014,drori_observation_2019}.
In the anomalous dispersion region, a soliton can be generated in the fibre \cite{Agrawal_2012}.
In this case, wavelengths in the visible/UV range might be velocity matched with the pulse: the emission from the front or back of the pulse will be found in this range, where the pulse edge constitutes a horizon.
Other nonlinear effects such as the Raman effect, phase-matched four-wave mixing or \v{C}erenkov radiation may take place in the medium but may be distinguished by their respective spectral signature.

\begin{table*}\setlength{\arrayrulewidth}{0.3mm}
\begin{center}
\begin{scriptsize}
	\begin{tabular}[t]{ |c|c|c|c|c|c|c|c|c|c|c|c| }
      \hline
      \hline
		\multicolumn{2}{|c|}{RIF}    &   \multicolumn{5}{c|}{White hole}   & \multicolumn{5}{c|}{Black hole}  			\tabularnewline
     		 \thead{$u/c$\\ \hfill} & \thead{$\lambda_c$\\(nm)} & \thead{$\lambda_{noL}$ \\ (nm)}& 		\thead{$\lambda_{loL}$ \\ ($\mu$m)}& \thead{$\Phi^{noL}(\lambda)$ \\ \hfill }& \thead{$\Phi^{loL}(\lambda)$ \\ \hfill}& \thead{C \\ \hfill}& \thead{$\lambda_{noL}$ \\ (nm)} &\thead{$\lambda_{moR}$ \\ (nm)}& \thead{$\Phi^{noL}(\lambda)$ \\ \hfill}& \thead{$\Phi^{moR}(\lambda)$ \\ \hfill}& \thead{C \\ \hfill}
             \tabularnewline 
             \hline
$2/3$ &  $400$& $227$ & $3.6$ & $3.5\times10^{13}$ & $7.8\times10^8$ & $0.92$ & $209.8$ & $398.5$ & $2.1\times10^{14}$ & $1.9\times10^{11}$ & $0.97$
			\tabularnewline
$2.04/3$ & $800$ & $379$ & $2.01$ & $3.4\times10^{13}$ & $1.5\times10^{10}$ & $0.99$ & $372$ & $810$ & $1.1\times10^{14}$ & $1.3\times10^{11}$ & $0.99$
			\tabularnewline
$2.05/3$ & $1260$ & $438$ & $1.36$ & $2.4\times10^{14}$ & $6.1\times10^{10}$ & $0.99$ & $438$ & $1317$ & $2.1\times10^{14}$ & $1.9\times10^{10}$ & $0.99$
			\tabularnewline
$2.04/3$  & $1990$ & $379$ & $2.01$ & $6.7\times10^{13} $ & $1.5\times10^{10}$ & $0.99$ & $372$ & $810$ & $1.2\times10^{14}$ & $1.4\times10^{11}$ & $0.99$
 			\tabularnewline 
			\hline
	\end{tabular}
\end{scriptsize}
\caption{Dependence of horizon physics on the velocity $u$ of the RIF. $\lambda_c$: effective RIF central wavelength.\label{tab:tableu}}
\end{center}
\end{table*} 

\section{Conclusion\label{sec:conclusion}}
We considered the mixing of modes of positive- and negative-norm in various kinematic scenarios in dispersive optical analogues to gravity.
The refractive index front (RIF) may act as a black- or white hole horizon or as a horizonless emitter simultaneously.
We presented a theoretical model (expanding on the canonical adaptation \cite{schutzhold_dielectricbh_2002,finazzi_quantum_2013} of the Hopfield model \cite{hopfield_theory_1958}), formulated observables in the relevant reference frames  and calculated the two-particle spectrum of an optical analogue for the first time.

We obtained a broad and structured spectrum with emission peaks in pairs of modes.
The emission is in entangled photon pairs and occurs strongly from horizons on a background of a broadband, weak and horizonless emission.
We change the height of the RIF over 5 orders of magnitude, and find that the signatures of horizon emission are not affected: the quantum correlation structure remains intact and the emission shape does not change.

These effects are largely robust against changes in dispersion as well, and we identify nonlinear waveguides as media of choice to observe horizon physics.
The resulting spectral peak, spectral correlations and their dependence on RIF velocity and medium dispersion are key identifiers of the type of spontaneous emission at the horizon by the Hawking effect or the horizonless emission.

This analytical method can be generalised to study other RIF profiles, such as a finite-length pulse, and to calculate the spectra and their spectral correlations.
This is essential to identify the optimal conditions to observe the Hawking effect.
Importantly, the method can be used to provide a theory trace for optical analogue experiments, to learn more about the Hawking effect, and to study a variety of effects such as cosmological pair creation by passing gravitational waves or expanding/contracting universes \cite{weinfurtner_cosmological_2009,eckel_expanding_2018,campo_inflationary_2005,campo_inflationary_2005-1,campo_inflationary_2006}, the so-called black hole laser \cite{gaona-reyes_laser_2017,bermudez_resonant_2019}, analogue wormholes, and the quasi-bound states of black holes \cite{patrick_black_2018}.

\begin{acknowledgments}
The authors are thankful to Bill Unruh, Joseph Cousins, Stephen Barnett, Vyome Singh and David Bermudez for insightful conversations. This work was funded by the EPSRC via Grant No. EP/L505079/1 and the IMPP.
\end{acknowledgments}


\appendix
\section{Field Theory\label{app:FT}}
\subsection{\label{subsec:FTIDD}Light in an inhomogeneous dispersive dielectric}
Following \cite{schutzhold_dielectricbh_2002,macher_black/white_2009,finazzi_quantum_2013} and \cite{Jacquet_quantum_2015}, we describe the interactions of light with an inhomogeneous and transparent dielectric by a microscopic model based on the Hopfield model \cite{hopfield_theory_1958}.
We consider one-dimensional scalar electromagnetic fields and operate at frequencies sufficiently far from the medium resonances to neglect absorption.
The medium consists of polarisable molecules --- oscillators with eigenfrequencies (resonant frequency) $\Omega_i$ and elastic constants $\kappa_i^{-1}$.
Since the wavelength of light is large compared to the molecular scale, we consider the dielectric in the continuum limit and describe the electric dipole displacement by the massive scalar field $P_i$.
The electromagnetic field (a massless scalar field) is described by $A$, the $x$-component of the vector potential $\vec{A}(X,T)$ (with $\vec{E}=-\partial_T \vec{A}$ in temporal gauge), where $X$ and $T$ are space and time in the laboratory frame.

Our study is based on the consideration of the step-like geometry of a RIF, that propagates at constant speed $u$ in the positive $X$-direction in the laboratory frame. 
The RIF is shown schematically in Fig.\ref{fig:HRstep} in the co-moving Lorentz frame coordinates $x$ and $t$.
We locate the boundary of the RIF at $x=0$. We focus entirely on the index change induced by the RIF, neglecting phonon interactions and phase-matched optical nonlinearities such as four-wave mixing.
In both homogeneous regions ($x \gtrless 0$), the interaction of the electromagnetic field with the three polarization fields of the medium is described by the Lagrangian density \cite{finazzi_quantum_2013,Jacquet_quantum_2015,hopfield_theory_1958}
\begin{equation}
\begin{split}
\label{eq:LagrangianHopfieldMF}
\mathcal{L}_{MF}&=\frac{(\partial_t A)^2}{8 \pi c^2}-\frac{(\partial_x A)^2}{8 \pi} +\sum_{i=1}^N\left(\frac{\gamma^2(\partial_t P_i-u\partial_x P_i)^2}{2\kappa_i\Omega_i^2}\right.\\ &\quad\left.-\frac{P_i^2}{2\kappa_i}+\frac{A\gamma}{c}(\partial_t P_i-u\partial_x P_i)\right),
\end{split}
\end{equation}
where $(\kappa_i\Omega_i^2)^{-1}$ is the inertia of oscillator $P_i$  and $\gamma=\left(1-u^2/c^2\right)^{-\sfrac{1}{2}}$.
Most transparent dielectrics are sufficiently well described by three resonances and so we use $N=3$.
The term linear in $A$ in Eq.\eqref{eq:LagrangianHopfieldMF} describes the coupling between the fields.
The Lagrangian density accounts for the free space and medium contributions to the field through the first two terms and the sum, respectively.
Dispersion enters as a time dependence of the addends of the summation.

By the principle of least action, we obtain the Hamiltonian density by varying the Lagrangian density \eqref{eq:LagrangianHopfieldMF} with respect to the canonical momentum densities of light and the polarisation fields.
From the Hamiltonian density follow the Hamilton equations, the equations of motion for the fields \cite{finazzi_quantum_2013,cohen-tannoudji_photons_2004}:
\begin{eqnarray}
\label{eq:fieldmotioneqMFFT}
\label{eq:fieldmotioneqMFFT1}
\dot{A}&=&4\pi c^2\Pi_A\\
\label{eq:fieldmotioneqMFFT2}
\dot{P}_i&=&
\frac{\kappa_i\Omega_i^2}{\gamma^2}
\left(\Pi_{P_i}-A\frac{\gamma}{c}\right)+uP_i^{'}\\
\label{eq:fieldmotioneqMFFT3}
\dot{\Pi}_A&=&\frac{A^{''}}{4\pi}+\sum_{i=1}^3\left(\frac{\kappa_i\Omega_i^2}{\gamma^2}
\left(\Pi_{P_i}-A\frac{\gamma}{c}\right)\right)\\
\label{eq:fieldmotioneqMFFT4}
\dot{\Pi}_{P_i}&=&-\frac{P_i}{\kappa_i}+u\Pi_{P_i}^{'},
\end{eqnarray}
where the derivatives are with respect to $(x, t)$ and $\Pi_i$'s denote the fields canonical conjugate to $A$ and $P_i$.
We go to Fourier space by $\partial_t \leftrightarrow -i\omega$ and $\partial_x \leftrightarrow ik$, where $k$ and $\omega$ are, respectively, the wavenumber and frequency in the moving frame. We obtain the generic Sellmeier dispersion relation of bulk transparent dielectrics \cite{Sellmeier_1871}, Eq.\eqref{eq:SellmDispRelMFbody} of the main text:
\begin{equation}
\label{eq:SellmDispRelMF}
c^2k^2=\omega^2+\sum_{i=1}^3\frac{4\pi\kappa_i\gamma^2\left(\omega+uk\right)^2}{1-\frac{\gamma^2\left(\omega+uk\right)^2}{\Omega_i^2}}.
\end{equation}
We complexify the massive field obtained from the action of \eqref{eq:LagrangianHopfieldMF} by identifying harmonic plane wave solutions  to \eqref{eq:fieldmotioneqMFFT}-\eqref{eq:fieldmotioneqMFFT4} of the form
\begin{equation}
\label{eq:PWsolsMF}
\vec{V}(x, t)=\vec{\bar{V}}({\omega})\,e^{ik x-i\omega t},
\end{equation}
where $\vec{V}$ is the eight-dimensional field vector $\vec{V}=(A\ P_1\ P_2\ P_3\ \Pi_A\ \Pi_{P_1}\ \Pi_{P_2}\ \Pi_{P_3})^T$.
We can denote a single frequency field vector as  $\vec{V}^{\alpha}({\omega})$, where $\alpha$ indicates a particular solution $k^\alpha$ of \eqref{eq:SellmDispRelMF} for $\omega$, \textit{i.e.}, a `mode'. 

By construction, the Lagrangian \eqref{eq:LagrangianHopfieldMF} is invariant under global phase shifts of the dynamic fields.
This continuous symmetry implies a conserved Noether current \cite{Jacquet_book_2018,cohen-tannoudji_photons_2004}.
As a result, the Klein-Gordon product
\begin{equation}
\label{eq:scalarproduct}
\left\langle \vec{V}_1,\vec{V}_2 \right\rangle =\frac{i}{\hbar}\int dx \, \vec{V}_1^\dagger(x,t)\,\left(\begin{array}{cc}0&\mathbb{1}_4\\-\mathbb{1}_4&0\end{array}\right)\, \vec{V}_2(x,t)
\end{equation}
is conserved, and so is the induced norm $< \vec{V},\vec{V} >^{1/2}$. Here $\mathbb{1}_4$ is the $4\times4$ identity matrix and the Planck constant prefactor was inserted for normalisation.
It can be shown that the Klein-Gordon norm of a positive (negative) laboratory frequency $\Omega$ field is also positive (negative) \cite{finazzi_quantum_2013,rubino_negative-frequency_2012,mclenaghan_compression_2014}.
As a result, waves of positive frequency $\omega$ in the moving frame can have either sign of the norm.
This is different from fluid systems, where the sign of the norm is equal to that of the wave number $K$.
In optical systems, the Hawking effect takes place in correlated photons of positive and negative frequency $\Omega$, not wavenumber.

We orthonormalise a set of single mode field vectors $\vec{V}^{\alpha}({\omega})$ ($\alpha=\alpha_1\,...\,\alpha_8$) using the condition \cite{finazzi_quantum_2013}:
\begin{equation}
\label{eq:normalisationcondition}
\left\langle \vec{V}({\omega})^{\alpha_1}, \vec{V}({\omega'})^{\alpha_2}\right\rangle=\sgn(\Omega)\,\delta_{\alpha_1\alpha_2}\, \delta(\omega-\omega').
\end{equation}
Here $\sgn$ is the sign function that determines whether the mode $\alpha_1$ has positive or negative norm. Moreover, by Lorentz transform $\Omega = \gamma\;(\omega+uk^{\alpha})$ and $K = \gamma\; (k^{\alpha}+u/c^2\,\omega)$.

We now turn to the non-uniform medium, which consists of two homogeneous regions separated by a RIF.
The index in each  homogeneous region is described by the dispersion relation \eqref{eq:SellmDispRelMF}, with dispersion parameters $\kappa_{i,R}$ ($\kappa_{i,L}$) and $\Omega_{i,R}$ ($\Omega_{i,L}$) in the right (left) region.
The index distribution in the moving frame is:
\begin{equation}
\label{eq:RIFheight}
n(x)=n_L \,\theta \left( -x \right) +n_R \,\theta \left( x \right) =n_R+\delta n\, \theta \left( -x \right).
\end{equation}
$\theta$ is the Heaviside step function and $n_R$ ($n_L$) is the index on the right (left) side.
In an extension of the oscillator model by P. Drude and H. A.  Lorentz, $\mu$ parametrises the change of dispersion constants that leads to the  index change $\delta n$ \cite{schubert_nonlinear_1986,Jacquet_book_2018}:
\begin{equation}
\label{eq:nlindex}
 \kappa_{i L}=\mu\kappa_{i R} \quad \quad \Omega_{i L}^2=\mu^{-1}\Omega_{i R}^2.
\end{equation} 
For small index changes it follows from \eqref{eq:SellmDispRelMF} that $\mu\approx1+2 (n_R -n_R^{-1})^{-1} \, \delta n$.

\begin{figure}[t]
\centering
\includegraphics[width=.85\columnwidth]{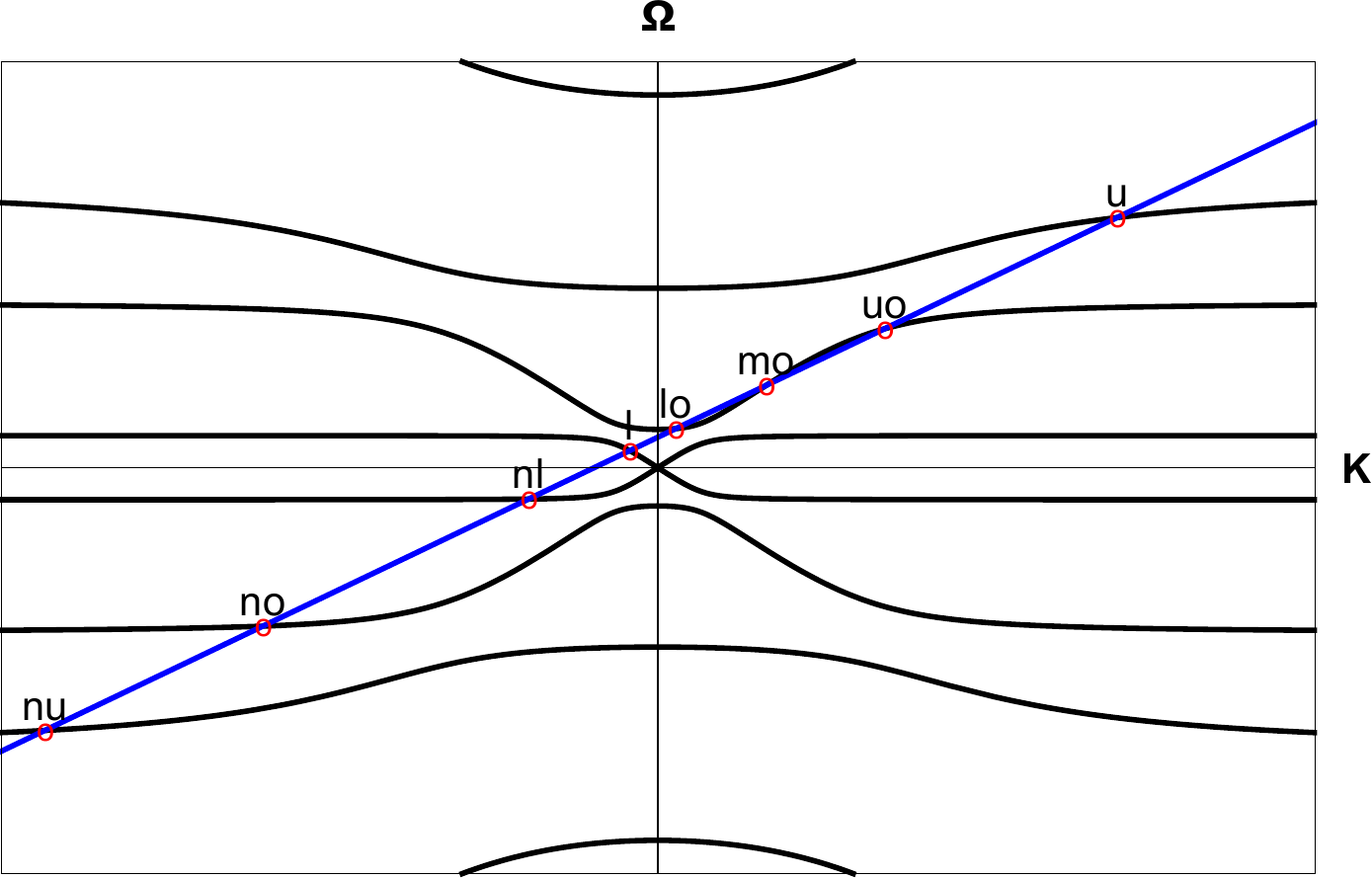}
\caption{Typical dispersion relation of a 3-resonances medium in the laboratory frame. There are eight branches (black curves). Plane wave mode solutions of \eqref{eq:SellmDispRelMF} are found at points of intersection with a contour of $\omega$ (straight blue line) and marked with red circles. \label{fig:labdisprel}}
\end{figure}

Harmonic wave solutions of frequency $\omega$ have a  propagation constant $k$ given by  Eq.\eqref{eq:SellmDispRelMF}, which is an eighth order polynomial, and thus eight wavenumbers $k^{\alpha}$ form the modes of the field $\vec{V}$ with degenerate energy $\hbar \omega$, as shown in Fig.\ref{fig:labdisprel}.
On either side of the RIF, there are either eight propagating modes or six propagating modes and two exponentially growing and decaying modes, respectively, characterized by complex $\omega$ and $k$.
The solution space is eight dimensional and we can build a full set of solutions for that space.
Modes can move towards and away from the RIF.

\subsection{Global modes of the system \label{subsec:GMsSM}}
In section \ref{sec:anagrav}, we have discussed all possible mode scenarios \cite{Jacquet_book_2018}. We proceed to construct modes of the inhomogeneous system, the  global modes (GMs).
These are solutions to the equation  of motion that are valid in both regions.
Global modes correspond to waves scattering at the RIF, and they describe the conversion of an incoming field to scattered fields in both regions.
The GMs are superpositions of the plane wave solutions in the two homogeneous regions on either side of the RIF.

We connect the plane wave solutions $\vec{V}(x, t)$,  the local modes (LMs), for the homogeneous medium at the  index boundary at $x=0$. 
We show in Appendix B, that the field $\vec{V}$ is continuous at the boundary.
The derivative of the field in space and time is also continuous, except for $\Pi_{P_i}$.
We refer to these relations as the `matching conditions'. 

We construct GMs $\mathcal{\vec{V}}$ as
\begin{equation}
\label{eq:GMdef}
\vec{\mathcal{V}}(x, t)=\sum_\alpha L^\alpha\, \vec{V}_L^\alpha(x, t)\,\theta(-x)+\sum_\alpha R^\alpha \,\vec{V}_R^\alpha(x, t)\,\theta(x),
\end{equation}
where $L^{\,\alpha}$ ($R^{\,\alpha}$) are coefficients of the eight modes $\alpha$ on the left (right) side of the RIF.
Because the fields and their conjugate momenta are related, half of the 16 coefficients in \eqref{eq:GMdef} are constrained by the matching conditions, leaving eight independent global modes.

\begin{figure}[t]
\centering
\includegraphics[width=\columnwidth]{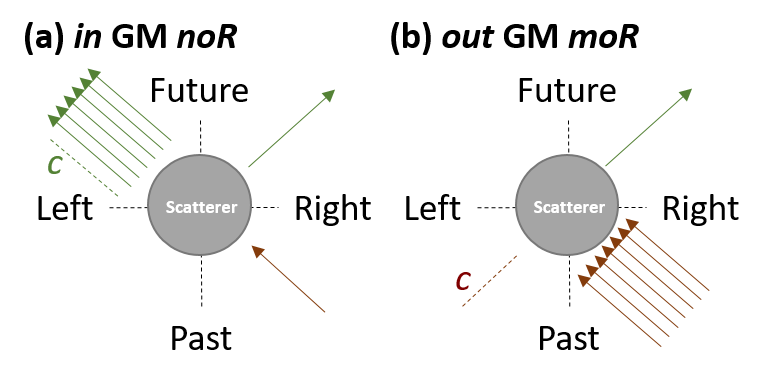}
\caption[Mode composition of the Gms]{Mode composition of two Global Modes (GMs) in a spacetime diagram: \textbf{(a)}, \textit{in} GM \textit{noR} --- there is a unique mode that propagates towards the RIF from the right (brown arrow). In the future, 6 oscillatory-modes propagate away from the RIF to the left, and one to the right, and there is one complex decaying mode (\textit{c}) on the left of the scatterer (RIF). \textbf{(b)}, \textit{out} GM \textit{moR} --- there is a unique mode that propagates away from the RIF to the right (green arrow). In the past, 7 oscillatory-modes propagate toward the RIF from the right and there is one complex decaying mode (\textit{c}) on the left of the scatterer.}\label{fig:modedecompomoR}
\end{figure}

Usually, the GMs are constructed as follows: a particular GM is constructed from a LM with its group velocity either towards (\textit{in}) or away from (\textit{out}) the RIF \cite{macher_black/white_2009}, irrespective of whether the LM is on the right or on the left.
A GM that emerges from a defining \textit{in} LM $\alpha$ forms a global \textit{in} mode $\vec{\mathcal{V}}^{in\,\alpha}$, see \textit{e.g.} Fig.\ref{fig:modedecompomoR} \textbf{a}.
\textit{out} LMs $\alpha$ define global \textit{out} modes $\vec{\mathcal{V}}^{out\,\alpha}$.
However, in order to accommodate for the unphysical modes, the procedure needs to be amended in the following way: if, on either side of the RIF, two modes are complex, they will have complex conjugate wave numbers and frequencies.
These are non-propagating solutions without group velocities and cannot be normalised.
In this case, the unbounded LM will define an unphysical GM, without other LMs on that side, which serves doubly as \textit{in} as well as \textit{out} mode.
This procedure always defines eight \textit{in} and \textit{out} GMs.
The LMs are complete solutions in the homogeneous regions, \textit{i.e.}, the sets $\vec{\mathcal{V}}^{in}=\{V^{in\;\alpha_1}...\,V^{in\;\alpha_8}\}$ and $\vec{\mathcal{V}}^{out}=\{V^{out\;\alpha_1}...\,V^{out\;\alpha_8}\}$ are two basis sets of the inhomogeneous medium.
Let us consider the example of the black hole-like case (mode scenario \textbf{d} in section \ref{sec:anagrav}). There is a unique \textit{out} GM, mode \textit{moR}, that allows for light to propagate away from the interface into the low index region.
Its mode decomposition is shown in a  spacetime diagram in Fig.\ref{fig:modedecompomoR} \textbf{b}: it is a linear combination of 7 oscillatory LMs, in the right region, that have negative group-velocity, a non-oscillatory (\textit{i.e.}, with complex wavenumber) LM on the left and a unique mode that has positive group-velocity in the right region.

\subsection{Quantum field theory of scattering at the RIF\label{subsec:QFTSM}}
Each \textit{in} GM describes the scattering of a harmonic wave to various outgoing harmonic waves.
Conversely, each \textit{out} GM describes a single harmonic wave resulting from the scattering of various incoming waves.
The scattering can be described in the \textit{in} as well as the \textit{out} basis.
The transformation between the two bases defines the scattering matrix, or $S$ matrix:
\begin{equation}
\label{eq:Smatrixdef}
\vec{\mathcal{V}}^{in\,\alpha}=\sum_\beta \vec{\mathcal{V}}^{out\,\beta}S_{\beta\,\alpha}.
\end{equation}
Forming a matrix $\mathcal{V}^{\mathrm{in}}$ ($\mathcal{V}^{\mathrm{out}}$) from the \textit{in} (\textit{out}) basis set, we describe the basis change as:
\begin{equation}
\begin{split}
\label{eq:fieldtransfo}
{\mathcal{V}}^\mathrm{in}&=\left(\vec{\mathcal{V}}^{\mathrm{in}\,\alpha_1}\ \vec{\mathcal{V}}^{\mathrm{in}\,\alpha_2}\ ... \ \vec{\mathcal{V}}^{\mathrm{in}\,\alpha_8}\right) \\ &=\left(\vec{\mathcal{V}}^{\mathrm{out}\,\alpha_1}\ \vec{\mathcal{V}}^{\mathrm{out}\,\alpha_2}\ ... \ \vec{\mathcal{V}}^{\mathrm{out}\,\alpha_8}\right)\, S={\mathcal{V}}^\mathrm{out}\,S.
\end{split}
\end{equation}

The spontaneous photon creation occurs because the quantum vacuum is basis dependent. Hence the spontaneous emission and all mode conversion follows from $S$.
We proceed with the canonical quantisation formalism introduced in \cite{hopfield_theory_1958}, developed in the 1990s in \cite{huttner_canonical_1991,huttner_dispersion_1992,matloob_electromagnetic_1995,barnett_spontaneous_1992,barnett_quantum_1995,santos_electromagnetic-field_1995}, and used in \cite{schutzhold_dielectricbh_2002,finazzi_quantum_2013,finazzi_spontaneous_2014,belgiorno_hawking_2015} and \cite{Jacquet_quantum_2015,Jacquet_book_2018} for the global modes.
We postulate the equivalent of the standard equal-time commutation relations on the fields $A$ and $P_i$:
\begin{equation}
\label{eq:commutA}
\left[A(x),\Pi_A(x')\right]=i\hbar\,\delta(x-x'),
\end{equation}
\begin{equation}
\label{eq:commutPi}
\left[P_i(x),\Pi_{P_j}(x')\right]=i\hbar\,\delta_{ij}\,\delta(x-x').
\end{equation}
We expand the quantised global field $\vec{\hat{\mathcal{V}}}$ in the basis of global \textit{in} modes:
\begin{equation}
\label{eq:GMinLMinbasis}
\begin{split}
\vec{\hat{\mathcal{V}}}&=\int\limits_0^\infty\mathrm{d\omega}\left( \sum_{\alpha\in P}\vec{\mathcal{V}}^{in\,\alpha}({\omega})\,\hat{a}^{in\,\alpha}({\omega})+\right.\\&\quad\left.\sum_{\alpha\in N}\vec{\mathcal{V}}^{in\,\alpha}({\omega})\,\hat{a}^{in\,\alpha\dagger}({\omega}) \right)+ \mathrm{H.c.},
\end{split}
\end{equation}
where P (N) denotes the set of all positive (negative) norm global modes (unphysical modes can be associated with positive norm here).
The expansion \eqref{eq:GMinLMinbasis} for \textit{in} (and its counterpart for \textit{out} modes) defines the annihilation operators  $\hat{a}^\alpha({\omega})$ and the creation operators $\hat{a}^{\alpha\dagger}({\omega})$ for each global mode $\alpha$, as well as the transformation between \textit{in} and \textit{out} creation and annihilation operators.
Hence, let $\vec{\hat{A}}^{in}$ be the column vector containing all the annihilation and creation operators for positive- and negative-norm global \textit{in} modes, respectively, and $\vec{\hat{A}}^{out}$ the corresponding vector for the \textit{out} modes. Then the transformation of operators follows from \eqref{eq:Smatrixdef} and \eqref{eq:GMinLMinbasis} as:
\begin{equation}
\label{eq:operatortransfo}
\vec{\hat{A}}^{out}=S\vec{\hat{A}}^{in}.
\end{equation}

With the scalar product \eqref{eq:scalarproduct} and the mode expansion \eqref{eq:GMinLMinbasis}  we can derive explicit expressions for the annihilation and creation operators and show that their commutator is:
\begin{equation}
\label{eq:commutator}
\left[\hat{a}^{\alpha}({\omega}),\hat{a}^{\alpha'\dagger}({\omega'})\right]= \delta_{\alpha \alpha'}\delta(\omega-\omega'),
\end{equation}
where the first $\delta$ is the Kronecker-delta. This relation holds for \textit{in} and \textit{out} operators. The commutator confirms that the global modes defined here are independent Bosonic modes in a continuous multimode field.

\section{Matching conditions of the fields at the RIF \label{subsec:matchingconditionsSM}}

Here we derive the continuity of the fields across the RIF. On physical grounds, we consider the field, polarisation field, all conjugate momenta and their time derivatives to be finite.
By construction of the model \cite{schubert_nonlinear_1986}, the elastic constant $\kappa$ is discontinuous  and the inertia of the polarisation fields $(\kappa \Omega^2)^{-1}$ is continuous at the interface between the two homogeneous regions (cf.   \eqref{eq:nlindex}). For fields of single frequency  $\omega$ the field's time derivatives do not alter the spatial continuity of the fields. 
In the near-interface region, we integrate the equations of motion over space.
We begin with \eqref{eq:fieldmotioneqMFFT3}: 
\begin{equation}
\label{eq:spaceintegralconjugatemomentumemfield}
\begin{split}
\int_{-\epsilon_1}^{+\epsilon_2}i\omega\,\Pi_A\mathrm{dx}&=\int_{-\epsilon_1}^{+\epsilon_2}\frac{A''}{4\pi}\mathrm{dx}+\\ &\quad\int_{-\epsilon_1}^{+\epsilon_2}\sum_{i=1}^3\frac{\kappa_i\Omega^2}{\gamma c}\left(\Pi_{P_i}-\gamma\frac
{A}{c}\right)\mathrm{dx}.
\end{split}
\end{equation}
All finite terms integrate to zero in the limits $(\epsilon_1\rightarrow0, \epsilon_2\rightarrow0)$ and so 
\begin{equation}
\label{eq:emfieldcontinuity}
\lim_{\substack{\epsilon_1 \to 0 \\ \epsilon_2 \to 0}}\int_{-\epsilon_1}^{+\epsilon_2}\frac{A''}{4\pi}\,\mathrm{dx}=0.
\end{equation}
Thus $A''$ is finite and the vector potential $A$ is continuously differentiable at $x=0$: $A (0^-,t)=A (0^+,t)$ and $A' (0^-,t)=A' (0^+,t)$.
Proceeding similarly with  \eqref{eq:fieldmotioneqMFFT2} leads to 
\begin{equation}
\label{eq:polfieldcontinuity}
\begin{split}
\lim_{\substack{\epsilon_1 \to 0 \\ \epsilon_2 \to 0}}\int_{-\epsilon_1}^{+\epsilon_2}P'_i\,\mathrm{dx}=0 \quad
\Rightarrow \quad P_{i}(0^-,t)=P_{i}(0^+,t),
\end{split}
\end{equation}
i.e. the polarisation fields are continuous at $x=0$. Analogously we find from   \eqref{eq:fieldmotioneqMFFT4} that the conjugate momenta $\Pi_{P_i}$'s are continuous.
We established now the continuity of $A$, $P_i$, $\Pi_{P_i}$, and $\kappa_i \Omega^2_i$  in  \eqref{eq:fieldmotioneqMFFT2}, and so $P'_i$, too, is continuous: $P_i' (0^-,t)=P_i' (0^+,t)$.
Finally, turning back to  \eqref{eq:fieldmotioneqMFFT4}, in which $P_i$ and $\dot{\Pi}_{P_i}$  are continuous, we see that discontinuity of $\kappa_i$ implies discontinuity of  $\Pi'_{P_i}$.
Subtracting the rhs of  \eqref{eq:fieldmotioneqMFFT4} on either side of the RIF at $x = 0$ we obtain:
\begin{equation}
\label{eq:discontinuityofspatialconjugatepolfield}
 \Pi'_{P_i }(0^+,t)=\Pi'_{P_i }(0^-,t)+\frac{P_i}{u}\left(\frac{1}{\kappa_{i,R}}-\frac{1}{\kappa_{i,L}}\right).
\end{equation}

\section{Scattering matrix from the matching conditions\label{subsec:smfrommcsSM}}
We explicitly derive the scattering matrix for the mode scenario of Fig.1 \textbf{c}.
This corresponds to a frequency $\omega$ with 8 propagating modes on either side of the RIF interface.
We use \eqref{eq:matchINLeftRightSigma} to obtain all $\sigma^{\mathrm{in}}$'s (or \eqref{eq:matchOUTLeftRightSigma} for $\sigma^{\mathrm{out}}$'s) from a single matrix
\begin{equation}
\label{eq:Amatrixdef}
\mathcal{A}=\left(W_L^{-1}\ W_R\right) \big|_{x=0}.
\end{equation}
Combining \eqref{eq:matchINLeftRightSigma} and \eqref{eq:Amatrixdef} we relate the $\sigma^{\mathrm{in}}$'s on either side of the RIF.
Each column of $\sigma^{\mathrm{in}}$ contains the coefficients of individual local modes to one of the global modes on that side of the RIF.
We arrange the global and local modes, respectively, in decreasing order of $\Omega$, i.e. first \textit{$u^{\mathrm{in}}$}, then \textit{$uo^{\mathrm{in}}$} and so on.
We use the construction of GMs described above eq.\eqref{eq:Amatrixdef}, with the modes ordered as $u,\; uo,\; mo,\;lo,\;l,\;nl,\;nol,\; nu$, and hence \eqref{eq:matchINLeftRightSigma} writes:
\begin{equation}
\label{eq:sigmamatricesoscosc}
\renewcommand*{\arraystretch}{0.9}
\left(\begin{array}{cccccccc}
\multicolumn{8}{c}{\hfill}\\[0.0em]\multicolumn{8}{c}{\hfill}\\[0.0em]0&0&1&0&0&0&0&0\\[0.0em] \multicolumn{8}{c}{\hfill} \\[0.0em] \multicolumn{8}{c}{\hfill} \\[0.0em] \multicolumn{8}{c}{\hfill} \\[0.0em] \multicolumn{8}{c}{\hfill} \\[0.0em] \multicolumn{8}{c}{\hfill}  \end{array}\right)=
\mathcal{A}
\left(\begin{array}{cccccccc}
1&0&0&0&0&0&0&0\\
0&1&0&0&0&0&0&0\\ \multicolumn{8}{c}{\hfill}
\\
0&0&0&1&0&0&0&0\\
0&0&0&0&1&0&0&0\\
0&0&0&0&0&1&0&0\\
0&0&0&0&0&0&1&0\\
0&0&0&0&0&0&0&1\end{array}\right)
\end{equation}
There are 64 unknowns, `empty' components of the matrices. We obtain:
\begin{widetext}
\begin{eqnarray}
\label{eq:sigmainL8x8}
\sigma_L^{in}&=&
\left(\begin{array}{ccccc}
\mathcal{A}_{11}-\frac{\mathcal{A}_{13}\mathcal{A}_{31}}{\mathcal{A}_{33}}&\mathcal{A}_{12}-\frac{\mathcal{A}_{13}\mathcal{A}_{32}}{\mathcal{A}_{33}}&\frac{\mathcal{A}_{13}}{\mathcal{A}_{33}}&\mathcal{A}_{14}-\frac{\mathcal{A}_{13}\mathcal{A}_{34}}{\mathcal{A}_{33}}&\cdots\\
\mathcal{A}_{21}-\frac{\mathcal{A}_{23}\mathcal{A}_{31}}{\mathcal{A}_{33}}&\mathcal{A}_{22}-\frac{\mathcal{A}_{23}\mathcal{A}_{32}}{\mathcal{A}_{33}}&\frac{\mathcal{A}_{23}}{\mathcal{A}_{33}}&\mathcal{A}_{24}-\frac{\mathcal{A}_{23}\mathcal{A}_{34}}{\mathcal{A}_{33}}&\cdots\\
0&0&1&0&\cdots\\
\mathcal{A}_{41}-\frac{\mathcal{A}_{43}\mathcal{A}_{31}}{\mathcal{A}_{33}}&\mathcal{A}_{42}-\frac{\mathcal{A}_{43}\mathcal{A}_{32}}{\mathcal{A}_{33}}&\frac{\mathcal{A}_{43}}{\mathcal{A}_{33}}&\mathcal{A}_{44}-\frac{\mathcal{A}_{43}\mathcal{A}_{34}}{\mathcal{A}_{33}}&\cdots\\
\vdots&\vdots&\vdots&\vdots&\ddots\\
\mathcal{A}_{81}-\frac{\mathcal{A}_{83}\mathcal{A}_{31}}{\mathcal{A}_{33}}&\mathcal{A}_{82}-\frac{\mathcal{A}_{83}\mathcal{A}_{32}}{\mathcal{A}_{33}}&\frac{\mathcal{A}_{83}}{\mathcal{A}_{33}}&\mathcal{A}_{84}-\frac{\mathcal{A}_{83}\mathcal{A}_{34}}{\mathcal{A}_{33}}&\cdots
\end{array}\right) \\ \vspace{10mm}
\label{eq:sigmainR8x8}
\sigma_R^{in}&=&
\left(\begin{array}{ccccc}
1&0&0&0&\cdots\\
0&1&0&0&\cdots\\
-\frac{\mathcal{A}_{31}}{\mathcal{A}_{33}}&-\frac{\mathcal{A}_{32}}{\mathcal{A}_{33}}&\frac{1}{\mathcal{A}_{33}}&-\frac{\mathcal{A}_{34}}{\mathcal{A}_{33}}&\cdots\\
0&0&0&1&\cdots\\
\vdots&\vdots&\vdots&\vdots&\ddots\\
0&0&0&0&\cdots
\end{array}\right).
\end{eqnarray}

For the \textit{out} modes, \eqref{eq:matchINLeftRightSigma} is
\begin{equation}
\label{eq:sigmamatricesoscoscout}
\renewcommand*{\arraystretch}{0.9}
\left(\begin{array}{cccccccc}
1&0&0&0&0&0&0&0\\
0&1&0&0&0&0&0&0\\ \multicolumn{8}{c}{\hfill}
\\
0&0&0&1&0&0&0&0\\
0&0&0&0&1&0&0&0\\
0&0&0&0&0&1&0&0\\
0&0&0&0&0&0&1&0\\
0&0&0&0&0&0&0&1\end{array}\right)
=\mathcal{A}\left(\begin{array}{cccccccc}
\multicolumn{8}{c}{\hfill}\\ \multicolumn{8}{c}{\hfill}\\0&0&1&0&0&0&0&0\\ \multicolumn{8}{c}{\hfill} \\  \multicolumn{8}{c}{\hfill}\\  \multicolumn{8}{c}{\hfill}\\  \multicolumn{8}{c}{\hfill}\\ \multicolumn{8}{c}{\hfill}\end{array}\right),
\end{equation}
and by comparison with \eqref{eq:sigmamatricesoscosc} we exchange L $\longleftrightarrow$ R and replace $\mathcal{A}$ by $\mathcal{A}^{-1}$ in \eqref{eq:sigmainL8x8} and \eqref{eq:sigmainR8x8}.  Furthermore, we invert $\sigma_L^{\mathrm{out}}$:
\begin{equation}
\label{eq:sigmaoutLTandTinv8x8}
{{\sigma_L^{out}}}^{-1}=\left(\begin{array}{ccccc}
1&0&0&0&\cdots\\
0&1&0&0&\cdots\\
\mathcal{A}_{31}^{-1}&\mathcal{A}_{32}^{-1}&\mathcal{A}_{33}^{-1}&\mathcal{A}_{34}^{-1}&\cdots\\
0&0&0&1&\cdots\\
\vdots&\vdots&\vdots&\vdots&\ddots\\
0&0&0&0&\cdots
\end{array}\right).
\end{equation}
Finally, by \eqref{eq:smatrixsigma}, we obtain the scattering matrix 
\begin{equation}
\label{eq:Smatrix8x8}
\begin{split}
S=\left(\begin{array}{ccccc}
\mathcal{A}_{11}-\frac{\mathcal{A}_{13}\mathcal{A}_{31}}{\mathcal{A}_{33}}&\mathcal{A}_{12}-\frac{\mathcal{A}_{13}\mathcal{A}_{32}}{\mathcal{A}_{33}}&-\frac{\mathcal{A}_{13}}{\mathcal{A}_{33}}&\mathcal{A}_{14}-\frac{\mathcal{A}_{13}\mathcal{A}_{34}}{\mathcal{A}_{33}}&\cdots\\
\mathcal{A}_{21}-\frac{\mathcal{A}_{23}\mathcal{A}_{31}}{\mathcal{A}_{33}}&\mathcal{A}_{22}-\frac{\mathcal{A}_{23}\mathcal{A}_{32}}{\mathcal{A}_{33}}&-\frac{\mathcal{A}_{23}}{\mathcal{A}_{33}}&\mathcal{A}_{24}-\frac{\mathcal{A}_{23}\mathcal{A}_{34}}{\mathcal{A}_{33}}&\cdots\\
-\frac{\mathcal{A}_{31}}{\mathcal{A}_{33}}&-\frac{\mathcal{A}_{32}}{\mathcal{A}_{33}}&\frac{1}{A_{33}}&-\frac{A_{34}}{\mathcal{A}_{33}}&\cdots\\
\mathcal{A}_{41}-\frac{\mathcal{A}_{43}\mathcal{A}_{31}}{\mathcal{A}_{33}}&\mathcal{A}_{42}-\frac{\mathcal{A}_{43}\mathcal{A}_{32}}{\mathcal{A}_{33}}&-\frac{\mathcal{A}_{43}}{\mathcal{A}_{33}}&\mathcal{A}_{44}-\frac{\mathcal{A}_{43}\mathcal{A}_{34}}{\mathcal{A}_{33}}&\cdots\\
\vdots&\vdots&\vdots&\vdots&\ddots\\
\mathcal{A}_{81}-\frac{\mathcal{A}_{83}\mathcal{A}_{31}}{\mathcal{A}_{33}}&\mathcal{A}_{82}-\frac{\mathcal{A}_{83}\mathcal{A}_{32}}{\mathcal{A}_{33}}&-\frac{\mathcal{A}_{83}}{\mathcal{A}_{33}}&\mathcal{A}_{84}-\frac{\mathcal{A}_{83}\mathcal{A}_{34}}{\mathcal{A}_{33}}&\cdots
\end{array}\right).
\end{split}
\end{equation}
\end{widetext}
In \eqref{eq:Smatrix8x8}, we have completed the derivation of the $S$ matrix for mode scenario \textbf{c} with 8 propagating modes on either side of the interface. 
The $\mathcal{A}$ coefficients are taken from \eqref{eq:Amatrixdef}, \textit{viz}. from the normalized local mode components.
This derivation follows on from the matching conditions for the fields and their first spatial derivative at the interface and results in a straightforward expression for $S$ that can easily be evaluated on a computer.

\section{Quasi-unitarity of the scattering matrix}\label{subsec:quasiunitaritySM}

The scattering matrix describes the basis change between  \textit{in}  and \textit{out} global modes. Both are orthonormal bases, at least when there are 8 propagating modes. In order to preserve orthonormality, the scattering matrix is constrained. Alternatively, this can be seen as the preservation of the commutator relation for \textit{in}  and \textit{out} annihilation and creation operators.

The orthonormality of the \textit{in} GMs $\alpha$ and $\alpha'$ is defined by a matrix $g$, 
\begin{equation}
\label{eq:gdef}
g_{\alpha\,\alpha'}=\left\langle \vec{\mathcal{V}}^{\mathrm{in}\,\alpha},\vec{\mathcal{V}}^{\mathrm{in}\,\alpha'}\right\rangle, 
\end{equation}
with respect to the scalar product \eqref{eq:scalarproduct}. The only non-zero elements  of $g$ are +1 or -1 on the diagonal, indicating the positivity or negativity of the mode norm. In the \textit{out} basis we have the same number of negative modes, because the norm is conserved during scattering, and so relation \eqref{eq:gdef} is also valid for \textit{out} GMs, if we order the modes accordingly. Using \eqref{eq:fieldtransfo} we calculate

\begin{equation}
\label{eq:derivequasiunitarity}
\begin{split}
g&=\frac{i}{\hbar}\int dx\, \mathcal{V}^{\mathrm{in}\, \dagger} \,\left(\begin{array}{cc}0&\mathbb{1}_4\\-\mathbb{1}_4&0\end{array}\right)\, \mathcal{V}^{\mathrm{in}} \\
&=\frac{i}{\hbar}\int dx\, S^{\dagger}\mathcal{V}^{\mathrm{out}\, \dagger} \,\left(\begin{array}{cc}0&\mathbb{1}_4\\-\mathbb{1}_4&0\end{array}\right)\,  \mathcal{V}^{\mathrm{out}}S\\
&=S^{\dagger}g\,S.
\end{split}
\end{equation}
This relation is called `quasi-unitarity' and means that $S$ (and $S^{\dagger}$) is a member of the indefinite unitary group $U(5,3)$. We can reformulate this condition as a normalization condition for the rows (and columns) of $S$ as:
\begin{equation}
\label{eq:Snorm}
\sgn({\Omega^{\alpha}})=\sum_{\alpha' \in P}\left|S_{\alpha \alpha'}(\omega)\right|^2-\sum_{\alpha' \in N}\left|S_{\alpha {\alpha'}}(\omega)\right|^2,
\end{equation}
where $\sgn$ indicates the frequency sign and thus the norm of mode $\alpha$.
Ensuring that the scattering matrix is quasi-unitary is a useful test for numerical implementations.
The procedure is easily generalised for other cases, where there are complex, non-propagating mode solutions. As these generate non-physical modes, the scattering matrix becomes a block matrix and, although not normalizable, the norm of the unphysical mode can be defined as unity in $g$.\\

\section{Higher order correlation function}\label{subsec:corrfunctionsSM}
Here we detail the calculation of photon number variances and covariances. These are expressed by the expectation value of the second and fourth order moments of the \textit{out} annihilation operators, which we calculate here. The expectation value is taken with respect to the \textit{in} vacuum state. Therefore, we write out the Bogoljubov transformation \eqref{eq:operatortransfo} \cite{Bogoljubov_1958}, which connects \textit{in} and \textit{out} operators, in the norm-independent way:
\begin{equation}
	\label{eq:Bogoljubov}
	\hat{a}^{\mathrm{out}\,\alpha}(\omega)=\!\!\!\!\sum\limits_{\beta \in \{ \alpha \}}\!\!\mathcal{S}_{\alpha\beta}(\omega)\, \hat{a}^{\mathrm{in}\,\beta}(\omega)+\!\!\!\!\sum\limits_{\beta \notin \{ \alpha \}}\!\!\mathcal{S}_{\alpha\beta}(\omega
) \,\hat{a}^{\mathrm{in}\,\beta\, \dagger}(\omega) ,
\end{equation}
where $\{ \alpha \}$ again stands for the set of modes with norm identical to $\alpha$. The matrix $\mathcal{S}$ is equal to the scattering matrix $S$ except for the rows which belong to negative norm modes, that are complex conjugated. Creation operators for the \textit{out} modes are then obtained by Hermitian conjugation of \eqref{eq:Bogoljubov} only. Note that this expression is valid for any mode $\alpha$, whether of positive or negative norm.

We start with the second moment 
\begin{align}
\label{eq:2ndmom1}
		&\langle 0^\mathrm{in}| \hat{a}^{\mathrm{out}\,\alpha\,\dagger}(\omega)\,\hat{a}^{\mathrm{out}\,\alpha'}(\omega') |0^\mathrm{in} \rangle \nonumber = \\
       \nonumber &\quad\sum\limits_{\substack{\beta, \beta' \notin \{ \alpha \},\{ \alpha' \}}}\!\!\!\!\!\!\! \mathcal{S}^*_{\alpha  \beta}(\omega) \,\mathcal{S}_{\alpha' \beta'}(\omega') \langle 0^\mathrm{in}| \hat{a}^{\mathrm{in}\,\beta}(\omega)\,\hat{a}^{\mathrm{in}\,\beta' \dagger}(\omega') |0^\mathrm{in} \rangle \\  & \quad=  \delta_{\{\alpha\}\{\alpha'\}}\, \delta(\omega-\omega')\,\sum\limits_{\beta \notin \{ \alpha \} } \mathcal{S}^*_{\alpha  \beta}(\omega) \,\mathcal{S}_{\alpha'  \beta}(\omega) .
\end{align}
In \eqref{eq:2ndmom1} we have used \eqref{eq:Bogoljubov} and that the annihilation operator applied to the vacuum vanishes. In the second step we also used the commutator \eqref{eq:commutator}. The spectral correlation is $\delta$-function peaked as expected for a stationary process; there are no positive-to-negative norm correlations in the fields.
From \eqref{eq:2ndmom1} and \eqref{eq:fluxop}, equation \eqref{eq:fluxbody} of the main text follows.
We now proceed to calculate higher order correlations.

In what follows we will drop the explicit \textit{in}-vacuum state in the expectation value and the upper index \textit{in} on the operators. We also leave out the frequency dependence of $\mathcal{S}$ ($\hat{a}$), as it corresponds with the first index of $\mathcal{S}$ (the mode of $\hat{a}$) in the moments calculation. Next, we calculate the normally ordered fourth order moment
\begin{widetext}
\begin{equation}
\label{eq:4thmom1}
\begin{split}
	\langle \hat{a}^{\mathrm{out}\,\alpha\,\dagger}\,\hat{a}^{\mathrm{out}\,\alpha'\,\dagger}\,\hat{a}^{\mathrm{out}\,\alpha''}\,\hat{a}^{\mathrm{out}\,\alpha'''} \rangle  &= \sum\limits_{\substack{ \beta, \beta''' \notin \{ \alpha \},\{ \alpha''' \} }} \!\!\!\!\!\!\mathcal{S}^*_{\alpha  \beta} \,\mathcal{S}_{\alpha''' \beta'''} \langle \hat{a}^{\beta}\,\hat{a}^{\mathrm{out}\,\alpha'\,\dagger}\,\hat{a}^{\mathrm{out}\,\alpha''}\,\hat{a}^{\beta''' \dagger} \rangle \\  &= \!\!\!\!\!\!\sum\limits_{\substack{ \beta, \beta'''  \notin \{ \alpha \},\{ \alpha''' \}\\ \beta', \beta'' \in \{ \alpha' \},\{ \alpha'' \} }}\!\!\!\!\!\! \mathcal{S}^*_{\alpha  \beta} \,\mathcal{S}^*_{\alpha'  \beta'} \,\mathcal{S}_{\alpha''  \beta''} \,\mathcal{S}_{\alpha''' \beta'''} \langle \hat{a}^{\beta}\,\hat{a}^{\beta'\,\dagger}\,\hat{a}^{\beta''}\,\hat{a}^{\beta''' \dagger} \rangle  \\ &\quad \quad+ \!\!\!\!\!\!\!\!\!\!\sum\limits_{\substack{ \beta, \beta' \notin \{ \alpha \},\{ \alpha' \}\\ \beta'', \beta''' \notin \{ \alpha'' \}, \{ \alpha''' \} }}\!\!\!\!\!\!\!\!\!\!\!\! \mathcal{S}^*_{\alpha  \beta} \,\mathcal{S}^*_{\alpha'  \beta'} \,\mathcal{S}_{\alpha''  \beta''} \,\mathcal{S}_{\alpha''' \beta'''} \langle \hat{a}^{\beta}\,\hat{a}^{\beta'}\,\hat{a}^{\beta''\,\dagger}\,\hat{a}^{\beta''' \dagger} \rangle,
\end{split}
\end{equation}
with steps analogous to \eqref{eq:2ndmom1} and realizing that expectation values with unequal numbers of annihilation and creation \textit{in}-operators vanish. Next,
\begin{equation}
\label{eq:moments}
\begin{split}
\!\!\!\!\langle \hat{a}^{\beta}\,\hat{a}^{\beta'\,\dagger}\,\hat{a}^{\beta''}\,\hat{a}^{\beta''' \dagger} \rangle&= \delta_{\beta \beta'} \, \delta_{\beta'' \beta'''} \,\delta(\omega-\omega')\,\delta(\omega''-\omega''')	\\ 
\!\!\!\!	\langle \hat{a}^{\beta}\,\hat{a}^{\beta'}\,\hat{a}^{\beta''\,\dagger}\,\hat{a}^{\beta''' \dagger} \rangle&= \delta_{\beta \beta''} \, \delta_{\beta' \beta'''} \,\delta(\omega-\omega'')\,\delta(\omega'-\omega''')\\&\quad+\delta_{\beta \beta'''} \, \delta_{\beta' \beta''} \,\delta(\omega-\omega''')\,\delta(\omega'-\omega''),
\end{split}
\end{equation}
due to the commutator. Inserting into \eqref{eq:4thmom1} and eliminating two sums with the Kronecker-deltas, we obtain the final expression
\begin{equation}
	\label{eq:4thmom2}
	\begin{split}
		\langle &\hat{a}^{\mathrm{out}\,\alpha\,\dagger} \,\hat{a}^{\mathrm{out}\,\alpha'\,\dagger}\,\hat{a}^{\mathrm{out}\,\alpha''}\,\hat{a}^{\mathrm{out}\,\alpha'''} \rangle \\
&= \delta_{ \{\alpha\}\overline{\{\alpha'\}}} \, \delta_{\overline{\{\alpha''\}} \{\alpha'''\}} \,\delta(\omega-\omega')\,\delta(\omega''-\omega''') \sum\limits_{\substack{ \beta, \beta'' \notin \{ \alpha \},\{ \alpha''' \} }} \mathcal{S}^*_{\alpha  \beta} \,\mathcal{S}^*_{\alpha'  \beta} \,\mathcal{S}_{\alpha''  \beta''} \,\mathcal{S}_{\alpha''' \beta''} \\ 
&\quad+\delta_{\{\alpha\} \{\alpha''\}} \, \delta_{\{\alpha'\} \{\alpha'''\}} \,\delta(\omega-\omega'')\,\delta(\omega'-\omega''')\sum\limits_{\substack{ \beta, \beta' \notin \{ \alpha \},  \{ \alpha''' \} }} \mathcal{S}^*_{\alpha  \beta} \,\mathcal{S}^*_{\alpha'  \beta'} \,\mathcal{S}_{\alpha''  \beta} \,\mathcal{S}_{\alpha''' \beta'} \\ 
&\quad+ \delta_{\{\alpha\} \{\alpha'''\}} \, \delta_{\{\alpha'\} \{\alpha''\}} \,\delta(\omega-\omega''')\,\delta(\omega'-\omega'')\sum\limits_{\substack{  \beta, \beta' \notin \{ \alpha \},  \{ \alpha' \} }} \mathcal{S}^*_{\alpha  \beta} \,\mathcal{S}^*_{\alpha'  \beta'}\,\mathcal{S}_{\alpha''  \beta'} \,\mathcal{S}_{\alpha''' \beta}.
	\end{split}
\end{equation}
	
In this expression we denote $\overline{\{\alpha\}}$ as the set of modes of norm opposite to that of mode $\alpha$. 
Finally, the \textit{not normally} ordered fourth order moment of mode $\alpha$ is
\begin{equation}
\label{eq:nnovariance}
\begin{split}
\langle  \hat{a}^{\mathrm{out}\,\alpha\,\dagger}\,\hat{a}^{\mathrm{out}\,\alpha}\,\hat{a}^{\mathrm{out}\,\alpha\,\dagger}\,\hat{a}^{\mathrm{out}\,\alpha} \rangle &= \langle  \hat{a}^{\mathrm{out}\,\alpha\,\dagger}\,\hat{a}^{\mathrm{out}\,\alpha\,\dagger}\,\hat{a}^{\mathrm{out}\,\alpha}\,\hat{a}^{\mathrm{out}\,\alpha} \rangle+\delta(\omega'-\omega'')\langle  \hat{a}^{\mathrm{out}\,\alpha\,\dagger}(\omega)\,\hat{a}^{\mathrm{out}\,\alpha}(\omega''') \rangle \\ &= \delta(\omega-\omega')\,\delta(\omega''-\omega''')\sum\limits_{\substack{ \beta \notin \{ \alpha \} }} |\mathcal{S}_{\alpha  \beta}(\omega)|^{2} \sum\limits_{\substack{ \beta \notin \{ \alpha \} }} |\mathcal{S}_{\alpha  \beta}(\omega'')|^2 \\ &\quad
+ \delta(\omega-\omega')\,\delta(\omega'-\omega'')\sum\limits_{\substack{ \beta \notin \{ \alpha \} }} |\mathcal{S}_{\alpha  \beta}(\omega)|^2 \sum\limits_{\substack{ \beta \notin \{ \alpha \} }} |\mathcal{S}_{\alpha  \beta}(\omega')|^2\\&\quad+ \delta(\omega-\omega''')\,  \delta(\omega'-\omega'')\sum\limits_{\beta \notin \{ \alpha \} } |\mathcal{S}_{\alpha  \beta}(\omega)|^2 ,
\end{split}
\end{equation}
\end{widetext}
which we obtain by applying the commutator \eqref{eq:commutator} in the first step and \eqref{eq:2ndmom1} and \eqref{eq:4thmom2} in the second step. The result leads to the variance of mode $\alpha$ \eqref{eq:variance}.
Note that, in arriving at this result, we have identified the detector frequency with a unique mode in general.
However, exceptions are possible: (a) The frequency separating two modes could lie inside the detected frequency interval. In this case the interval is reduced to the mode with group velocity away from the RIF. (b) Two out-LMs from either side of the RIF might share a laboratory frequency $\Omega$. In this case they typically do not share the moving frame frequency $\omega$ and the covariances and variances of the modes add up.
\end{document}